\documentclass[reqno,12pt,dvips]{amsart}

\usepackage{amssymb,eucal}
\usepackage{array}
\usepackage{times,mathptmx}
\usepackage{psfrag}
\usepackage[dvips]{graphicx}
\usepackage{cite}
\usepackage{enumitem}
\usepackage{setspace}
\usepackage{geometry}

\numberwithin{equation}{section}

\allowdisplaybreaks

\newcommand{\group}[1]{\mathsf{#1}}
\newcommand{\alg}[1]{\mathfrak{#1}}

\newcommand{\func}[2]{#1 \left( #2 \right)}

\newcommand{\uealg}[1]{\mathcal{U} \bigl( #1 \bigr)}

\newcommand{\brac}[1]{\left( #1 \right)}
\newcommand{\sqbrac}[1]{\left[ #1 \right]}

\newcommand{\set}[1]{\left\{ #1 \right\}}
\newcommand{\pair}[2]{\left( #1 , #2 \right)}

\newcommand{\abs}[1]{\left| #1 \right|}

\newcommand{\ZZ}{\mathbb{Z}}

\newcommand{\CC}{\mathbb{C}}

\newcommand{\dd}{\mathrm{d}}
\newcommand{\ii}{\mathfrak{i}}

\newcommand{\eps}{\varepsilon}

\newcommand{\comm}[2]{\bigl[ #1 , #2 \bigr]}
\newcommand{\acomm}[2]{\bigl\{ #1 , #2 \bigr\}}
\newcommand{\dcomm}[2]{\bigl[ \mspace{-4.3mu} \bigl[ #1 , #2 \bigr] \mspace{-4.3mu} \bigr]}

\newcommand{\bra}[1]{\bigl\langle #1 \bigr\rvert}
\newcommand{\ket}[1]{\bigl\lvert #1 \bigr\rangle}
\newcommand{\braket}[2]{\bigl\langle #1 \bigr\rvert \bigl. #2 \bigr\rangle}
\newcommand{\bracket}[3]{\bigl\langle #1 \bigr\rvert #2 \bigl\lvert #3 \bigr\rangle} 

\newcommand{\radord}[1]{#1}
\newcommand{\normord}[1]{{} : #1 : {}} 

\newcommand{\affine}[1]{\widehat{#1}}
\newcommand{\killing}[2]{\kappa \bigl( #1 , #2 \bigr)}
\newcommand{\otherkilling}[2]{\widetilde{\kappa} \bigl( #1 , #2 \bigr)}
\newcommand{\strconst}[3]{f^{#1 #2}_{\phantom{#1 #2} #3}}

\newcommand{\gcreq}[1]{\overset{ #1 }{=}}

\newcommand{\MinMod}[2]{\mathcal{M} \left( #1 , #2 \right)}

\newcommand{\qfact}[2]{\left( #1 \right)_{#2}}

\newcommand{\rplus}{\mspace{4mu} \raisebox{-0.11em}{\reflectbox{\includegraphics[width=0.7em]{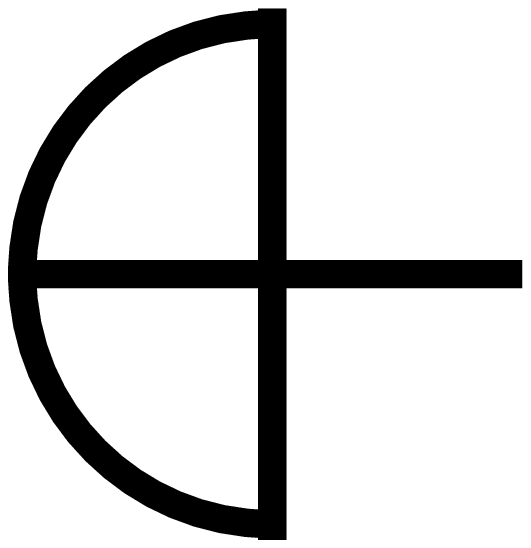}}} \mspace{4mu}}

\newcommand{\eqnref}[1]{Equation~(\ref{#1})}
\newcommand{\eqnDref}[2]{Equations~(\ref{#1}) and (\ref{#2})}

\newcommand{\secref}[1]{Section~\ref{#1}}
\newcommand{\secDref}[2]{Sections~\ref{#1} and \ref{#2}}
\newcommand{\figref}[1]{Figure~\ref{#1}}

\newcommand{\propref}[1]{Proposition~\ref{#1}}
\newcommand{\lemref}[1]{Lemma~\ref{#1}}
\newcommand{\thmref}[1]{Theorem~\ref{#1}}
\newcommand{\corref}[1]{Corollary~\ref{#1}}

\newcommand{\WZW}{Wess-Zumino-Witten}
\newcommand{\cft}{conformal field theory}
\newcommand{\cfts}{conformal field theories}
\newcommand{\ope}{operator product expansion}
\newcommand{\opes}{operator product expansions}
\newcommand{\ihw}{integrable highest weight}
\newcommand{\hws}{highest weight state}
\newcommand{\hwss}{highest weight states}
\newcommand{\hwm}{highest weight module}
\newcommand{\hwms}{highest weight modules}
\newcommand{\uea}{universal enveloping algebra}
\newcommand{\gcr}{generalised commutation relation}
\newcommand{\gcrs}{generalised commutation relations}

\DeclareMathOperator{\id}{id}
\DeclareMathOperator{\tr}{tr}
\DeclareMathOperator{\ad}{ad}
\DeclareMathOperator{\vectspan}{span}

\newtheorem{theorem}{Theorem}[section]				
\newtheorem{lemma}[theorem]{Lemma}					
\newtheorem{proposition}[theorem]{Proposition}	
\newtheorem{corollary}[theorem]{Corollary}		

\begin{document}

\title[Extended Algebras of $\func{\group{SU}}{2}$ WZW Models]{The Extended Algebra of the $\func{\group{SU}}{2}$ Wess-Zumino-Witten Models}

\author[P Mathieu]{Pierre Mathieu}

\address[Pierre Mathieu]{
D\'{e}partement de Physique, de G\'{e}nie Physique et d'Optique \\
Universit\'{e} Laval \\
Qu\'{e}bec, Canada G1K 7P4
}

\email{pmathieu@phy.ulaval.ca}

\author[D Ridout]{David Ridout}

\address[David Ridout]{
D\'{e}partement de Physique, de G\'{e}nie Physique et d'Optique \\
Universit\'{e} Laval \\
Qu\'{e}bec, Canada G1K 7P4
}

\email{darid@phy.ulaval.ca}

\thanks{\today \\ This work is supported by NSERC}

\begin{abstract}
The \WZW{} model defined on the group $\func{\group{SU}}{2}$ has a unique (non-trivial) simple current of conformal dimension $k/4$ for each level $k$.  The extended algebra defined by this simple current is carefully constructed in terms of generalised commutation relations, and the corresponding representation theory is investigated.  This extended algebra approach is proven to realise a faithful (``free-field-type'') representation of the $\func{\group{SU}}{2}$ model.  Subtleties in the formulation of the extended theory are illustrated throughout by the $k=1$, $2$ and $4$ models.  For the first two cases, bases for the modules of the extended theory are given and rigorously justified.
\end{abstract}

\maketitle

\onehalfspacing

\section{Introduction} \label{secIntro}

\subsection{Motivation and Outline} \label{secMotivOutline}

This work initiates a series of articles devoted to the study of conformal field theories reformulated in terms of an extended algebra, in which the algebra is augmented by a simple current.

The extension studied here is that of the diagonal \WZW{} models defined on $\func{\group{SU}}{2}$ \cite{KniCur84,WitNon84}.  These models are unitary, rational \cfts{} parametrised by a non-negative integer $k$ called the level.  The irreducible (unitarisable) representations from which the (chiral) \cft{} is constructed are the level-$k$ \ihw{} modules of $\func{\affine{\alg{sl}}}{2}$ \cite{GepStr86,FelSpe88}, of which there are $k+1$.  We denote the corresponding \hwss{} by $\ket{\psi_{\lambda}}$, for $\lambda = 0, 1, \ldots, k$, so $\lambda$ labels the corresponding $\func{\alg{sl}}{2}$-weight.  Among the corresponding primary fields $\func{\psi_{\lambda}}{z}$, there is a unique (non-trivial) simple current $\phi \equiv \psi_k$ (when $k \neq 0$), whose fusion rules take the form \cite{DiFCon97}
\begin{equation}
\phi \times \psi_{\lambda} = \psi_{k - \lambda}.
\end{equation}
This simple current therefore acts as a permutation on the irreducible modules of the chiral theory.  In particular, $\phi \times \phi = \psi_0$, the identity.

The construction considered here consists in augmenting the $\func{\affine{\alg{sl}}}{2}$ symmetry algebra of the theory by including the modes of the simple current.  It is actually convenient to consider the corresponding multiplet of \emph{zero-grade} descendant fields, rather than just the affine primary, whose corresponding states form an irreducible representation of $\func{\alg{sl}}{2}$.  We will therefore enlarge our algebra to include the modes of each zero-grade field $\func{\phi^{\brac{n}}}{z} $,  $n = 0, 1, \ldots, k$.  Because these fields each have conformal dimension $k/4$ (see \eqnref{eqnHWSConfDim}), this augmentation will in general lead to an extended algebra defined by \gcrs{}.  Our aim is to precisely define the structure of this extended algebra.

We note that such an investigation of this extended symmetry algebra has already been carried out for $k=1$ \cite{BerSpi94,BouSpi94}, under the guise of spinon bases.  For higher values of $k$, our description differs from the spinon formalism presented in \cite{BouSpi95, FeiMon05}, in that our algebra is generated by the simple current $\phi = \psi_{k}$ rather than $\psi_{1}$, whose fusion rules with the different primary fields are more complicated than a simple permutation.  We present here a formalism which applies to the $\func{\group{SU}}{2}$ models for all $k$, providing a detailed model for similar treatments of more general extended symmetries.  

Reformulations of \cfts{} of the type considered here are of particular interest in that they embody a quasi-particle description of the space of states, that is, the states are generated by the repeated action of the simple current modes.  This action is subject to restriction rules that can be interpreted as a generalised exclusion principle.  Such quasi-particle reformulations have a number of immediate implications, the most obvious being that they shed new light on the representation theory of the class of models under consideration.  In the particular case of the \WZW{} models, this translates into a novel point of view on the representation theory of affine Lie algebras.

The key point underlying this quasi-particle reformulation is the absence of singular vectors, meaning that the restriction rules are built into the algebraic structure (the \opes{}) of the extended theory.  Establishing this very point in the present context is our main result.  It follows that whenever the generating function for the quasi-particles can be found in a closed form, it will take the form of a positive-definite multiple sum, yielding a so-called \emph{fermionic} character formula.  We will derive such character formulae in several simple cases.  This  contrasts with the usual formulae deduced via the subtraction of singular vectors, leading to the non-positive-definite sums usually referred to as \emph{bosonic} character formulae.  Equating the fermionic and bosonic characters therefore results in non-trivial $q$-series identities that are of much combinatorial interest.

These two motivations are clearly rather mathematical in nature.  On a more physical note, we stress that quasi-particle formulations share the underlying spirit of the condensed-matter/particle formalism of \cft{}, rather than that of the algebraic formalism focussing largely on the representation theory of the chiral algebra.  The quasi-particle interpretation is therefore bound to be better adapted to the study of off-critical perturbations of the theory.  In that vein, a precise objective is to formulate the problem of computing correlation functions in a way that is suited to the faithful (free of singular vectors) description we construct here.  Moreover, we expect that it could provide a starting point for the evaluation of off-critical correlation functions.  Along integrable lines, these should be governed by integrable differential equations.  In brief, we expect this formulation to allow us to cross the free-fermion barrier in the description of correlation functions as solutions of solitonic equations, and hope to report on this elsewhere.

The extended algebra approach that we outline here was inspired by \cite{JacQua06, JacEmb06}, which present a description of the $\MinMod{3}{p}$ minimal models from the point of view of the extended algebra generated by $\phi_{2,1}$.  Its extension to all minimal models will be reported in \cite{RidMin06}.

\subsection{Organisation of the Article} \label{secOrg}

The construction of the extended algebra of the $\func{\group{SU}}{2}$ \WZW{} models is presented in \secref{secExtAlg}.  As mentioned above, this algebra combines the usual affine currents with the $k+1$ zero-grade components of the simple current.  The first issue we address is the question of the mutual locality of these two types of currents.  Overlooking this point quickly leads to subtle inconsistencies in the extended theory. It turns out that these considerations, and in particular imposing a natural definition for the adjoint of the simple current modes, force (somewhat surprisingly) the components of this simple current to \emph{anticommute} with certain of the affine generators.  This is shown to be compatible with the Jacobi identities and forms the conclusion of Section \ref{secAlgPrelim}.

The extended algebra is essentially defined by the \opes{} $\phi^{(m)}(z)\phi^{(n)}(w)$ of the simple current component fields, as the other generators, the affine currents, will be recovered (roughly speaking) as bilinears in these fields.  The first step towards computing the different fields appearing in these expansions is to consider the corresponding states.  These are computed for the first few levels, and are expressed directly in terms of the modes of the affine generators in \secref{secAlgOPEs}.  We then consider the constraints imposed by associativity in Section \ref{secAlgAssoc}.  The key point here is to show that the fields of our \opes{} can be constructed from these states in the usual way (the state-field correspondence).

More precisely, we have to address the following technical issue.  In the analysis of the $\MinMod{3}{p}$ models in \cite{JacQua06}, it was observed that the \ope{} corresponding to the fusion rule $\phi_{2,1} \times \phi_{2,1} = I$ must necessarily involve an operator $\tilde{\mathcal{S}}$ satisfying $\tilde{\mathcal{S}} \phi_{2,1} = (-1)^p \phi_{2,1} \tilde{\mathcal{S}}$.  The question of the potential necessity of a similar operator in the present context is addressed here, but is answered in the negative.  Nevertheless, we find it convenient to introduce trivial $\tilde{\mathcal{S}}$-type operators, denoted by $\mathcal{S}_{m,n}$ into each of our \opes{}, whose eigenvalues keep track of various constants that appear in the extended theory.  The trivial nature of these operators reflects the fact that they commute with the modes of the current generators and those of the simple current components.  With these technicalities settled, the path to the formulation of the \gcrs{} defining the extended algebra is completely paved.  These relations are displayed in \secref{secAlgGCRs}.  

The representation theory of this extended algebra is developed in \secref{secRepTheory}.  Previously, in order to write down the  \gcrs{}, we expanded the simple current components into modes acting on the vacuum module.  To formulate this expansion in full generality, we introduce the monodromy charge (\secref{secRepCharges}).  With this tool in hand, the notion of \hwss{} for the extended algebra is defined (Section \ref{secRepHWMs}), and the Verma modules for the extended algebra are constructed from the action of the simple current modes on such a state.  Within each extended Verma module, we readily identify two affine highest weight states.  Finally, we evaluate the eigenvalues (necessary for computation) of the ${\mathcal{S}}_{m,n}$ on these Verma modules in \secref{secSEigs}.

We then illustrate the general theory we have developed with two examples, $k=2$ and $k=4$, worked out in detail (\secref{secExamples}).  When $k=2$, the dimension of the three components of the simple current is $\frac{1}{2}$, which suggests a fermionic interpretation. Linear combinations of these three components are readily seen to correspond to three independent free fermions.  It is actually well-known that the affine $\func{\affine{\alg{sl}}}{2}_2$ algebra can be represented in terms of three free fermions \cite{GodKac85} (see also \cite[Sec.\ 15.5.5 and Ex.\ 15.16]{DiFCon97}).  We, however, show something more precise:  The $k=2$ extended algebra is \emph{isomorphic} to the tensor product of three copies of the extended symmetry algebra of the Ising model (defined by its own simple current), which is the symmetry algebra of a free fermion.  As the spectra (the highest-weight states) of the two theories also match, this suggests an equivalence of \cfts{}.

Another special case of interest is $k=4$, for which the extended algebra is shown to be isomorphic to the (affine) symmetry algebra of the $\func{\group{SU}}{3}$ \WZW{} model at level $1$, sharpening the correspondence $\func{\affine{\alg{sl}}}{2}_4 \sim \func{\affine{\alg{sl}}}{3}_1$ \cite{BaiAcc86} (see also \cite[Sec.\ 17.5.2]{DiFCon97}).  However, the comparison of the spectra is more delicate in this case, because we must take the non-diagonal $\func{\affine{\alg{sl}}}{2}_4$ model in this correspondence.  We show that the monodromy charge automatically accounts for this subtlety in our formalism.

The central result of this paper is presented in \secref{secSingVects}.  It states that the representations of the extended algebras we have constructed are faithful.  In other words, the Verma modules have no singular vectors, hence are irreducible. This result is established in various steps. First, we demonstrate that not only are there two affine highest-weight states in each extended Verma module, but that there are no more than two. For $k=1$ and $k=2$, we then show explicitly that there are no singular vectors in the (extended) Verma modules by demonstrating that the primitive singular vectors of the affine algebra vanish identically.  These calculations then motivate the general argument establishing the absence of singular vectors as presented in \secref{secSingVectsGenVan}.  This allows us to strengthen our conclusions concerning the cases $k=2$, $4$ and their connection to the theory of three fermions and the $\func{\affine{\alg{sl}}}{3}_1$ model respectively.  The isomorphisms, previously established for the symmetry algebras, are now extended to isomorphisms of irreducible modules.

Finally, a basis of states is rigorously derived for the two simplest models, $k=1$ and $2$.  The generating functions for these bases are constructed explicitly and shown to agree with known fermionic expressions for the corresponding characters.  We defer the derivation of the analogous basis of states for general $k$ to a sequel \cite{RidSU207}.

\subsection{Basics and Notation} \label{secBasics}

The $\func{\group{SU}}{2}$ models are unitary, rational \cfts{} parametrised by a non-negative integer $k$ called the level.  The chiral algebra of these models may be described as follows.  We first define \cite{KacInf90} the (untwisted) affine Lie algebra $\func{\affine{\alg{sl}}}{2}$ associated to the complexification $\func{\alg{sl}}{2}$ of $\func{\alg{su}}{2}$.  This is the (graded) complex Lie algebra
\begin{equation}
\func{\affine{\alg{sl}}}{2} = \func{\alg{sl}}{2} \otimes \CC \sqbrac{t , t^{-1}} \oplus \vectspan_{\CC} \set{K} \rplus \vectspan_{\CC} \set{L_0}
\end{equation}
($\rplus$ denotes a semidirect sum) whose non-trivial commutation relations are
\begin{equation} \label{eqnAffCommRels}
\begin{split}
\comm{L_0}{J \otimes t^m} &= -m J \otimes t^m \\
\text{and} \qquad \comm{J^1 \otimes t^m}{J^2 \otimes t^n} &= \comm{J^1}{J^2} \otimes t^{m+n} + m \killing{J^1}{J^2} \delta_{m+n,0} K.
\end{split}
\end{equation}
Note that $K$ is central.  Throughout we will follow the standard practice of abbreviating $J \otimes t^n$ as $J_n$.  Here, $\killing{J^1}{J^2}$ denotes the Killing form of $\func{\alg{sl}}{2}$, which we normalise so that
\begin{equation}
\killing{J^1}{J^2} = \frac{1}{4} \tr \Bigl[ \func{\ad}{J^1} \func{\ad}{J^2} \Bigr],
\end{equation}
where $\func{\ad}{\cdot}$ is the usual adjoint action of $\func{\alg{sl}}{2}$ on itself.  The chiral algebra is obtained from the \uea{} of $\func{\affine{\alg{sl}}}{2}$ by dropping $L_0$ (it will shortly be recovered through the Sugawara construction) and restricting $K$ to have eigenvalue $k$.  Formally, we take the \uea{} of the derived subalgebra of $\func{\affine{\alg{sl}}}{2}$ and form the quotient by the ideal generated by $K - k \id$.  We denote this chiral algebra by\footnote{We remark that it is far more commonplace in the physics literature to denote this chiral algebra by $\func{\affine{\alg{su}}}{2}_k$.  Whilst the name we ascribe an object is largely arbitrary, there is much to be said for maintaining consistent notation within a field.  That said however, we prefer to use the more precise notation given above as in the mathematics literature.  The reason why we explicitly indicate the \uea{} is that our chief object of study is a certain extension of this chiral algebra, and this extension is \emph{not} a Lie algebra.  We want to treat these extended algebras on an equal footing with more familiar chiral algebras, and so we emphasise chiral algebras as associative algebras, not Lie algebras.  The reason for writing $\alg{sl}$ rather than $\alg{su}$ is really a matter of taste --- $\func{\alg{su}}{2}$ does not, strictly speaking, contain the root vectors or coroot that form the most convenient basis for our study.} $\uealg{\func{\affine{\alg{sl}}}{2}_k}$.

We denote the elements of this basis by $\set{E, H, F}$, so that
\begin{equation} \label{eqnSL2Basis}
\comm{H}{E} = 2 E, \qquad \comm{E}{F} = H, \qquad \text{and} \qquad \comm{F}{H} = 2 F.
\end{equation}
The normalisation of the Killing form then gives
\begin{equation}
\killing{E}{E} = \killing{E}{H} = \killing{H}{F} = \killing{F}{F} = 0, \quad \killing{E}{F} = 1, \quad \text{and} \quad \killing{H}{H} = 2.
\end{equation}
The corresponding affine modes are therefore denoted by $E_n$, $H_n$ and $F_n$.

For a given level $k$, the Virasoro symmetry is realised as a subalgebra $\alg{Vir}_c$ of the chiral algebra through the Sugawara construction (see for example \cite[Eq.\ 15.70]{DiFCon97}):
\begin{equation} \label{eqnSugawara}
L_n = \frac{1}{2 \brac{k+2}} \sum_{m \in \ZZ} \normord{\frac{1}{2} H_m H_{n-m} + E_m F_{n-m} + F_m E_{n-m}}.
\end{equation}
Through this construction, the $L_0$ of \eqnref{eqnAffCommRels} is recovered, and is associated to the Virasoro energy operator.  The central charge of our \cft{} is then found to be
\begin{equation}
c = \frac{3 k}{k+2}.
\end{equation}
As previously indicated, the \hwss{} are denoted by $\ket{\psi_{\lambda}}$, for $\lambda = 0, 1, \ldots, k$, where $\lambda$ labels the $\func{\alg{sl}}{2}$-weight.  The conformal dimensions of these states are given by
\begin{equation} \label{eqnHWSConfDim}
h_{\lambda} = \frac{\lambda \brac{\lambda + 2}}{4 \brac{k+2}}.
\end{equation}
$\ket{\psi_0}$ is the vacuum of the theory, and will be denoted in what follows by $\ket{0}$.

We choose a basis of these zero-grade descendant states carefully in order to simplify later calculations.  First, we normalise the \hwss{}
\begin{equation}
\ket{\psi_{\lambda}} \equiv \ket{\psi_{\lambda}^{\brac{0}}}
\end{equation}
to have length $1$, and then define
\begin{equation}
\ket{\psi_{\lambda}^{\brac{n+1}}} = \bigl[ \brac{n+1} \brac{\lambda - n} \bigr]^{-1/2} F_0 \ket{\psi_{\lambda}^{\brac{n}}},
\end{equation}
for all $n = 0, 1, \ldots, \lambda - 1$.  The $\func{\alg{sl}}{2}$-weight of $\ket{\psi_{\lambda}^{\brac{n}}}$ is therefore $\lambda - 2n$:
\begin{equation}
H_0 \ket{\psi_{\lambda}^{\brac{n}}} = \brac{\lambda - 2n} \ket{\psi_{\lambda}^{\brac{n}}}.
\end{equation}
A simple induction argument now shows that
\begin{equation} \label{eqnSU2Normalisation}
F_0 \ket{\psi_{\lambda}^{\brac{n}}} = \bigl[ \brac{n+1} \brac{\lambda - n} \bigr]^{1/2} \ket{\psi_{\lambda}^{\brac{n+1}}} \quad \Rightarrow \quad E_0\ket{\psi_{\lambda}^{\brac{n+1}}} = \bigl[ \brac{n+1} \brac{\lambda - n} \bigr]^{1/2} \ket{\psi_{\lambda}^{\brac{n}}}.
\end{equation}
It follows from these definitions that the zero-grade descendant states are all normalised:
\begin{equation}
\braket{\psi_{\lambda}^{\brac{n+1}}}{\psi_{\lambda}^{\brac{n+1}}} = \bigl[ \brac{n+1} \brac{\lambda - n} \bigr]^{-1/2} \bracket{\psi_{\lambda}^{\brac{n+1}}}{F_0}{\psi_{\lambda}^{\brac{n}}} = \braket{\psi_{\lambda}^{\brac{n}}}{\psi_{\lambda}^{\brac{n}}},
\end{equation}
since $F_0^{\dag} = E_0$.

\section{The Extended Algebra} \label{secExtAlg}

Before starting the analysis of the extended algebra \emph{per se}, we first need to clear up a little technical issue, namely the analysis of the mutual locality of the two families of fields that we wish to combine into an extended chiral algebra\footnote{Chiral algebras are usually defined in relation with modular invariants and pertain to algebras spanned by fields with integer dimension.  Here we understand chiral algebra in a generalised sense, without reference to a modular invariant and thus without this dimensional  restriction.}.

\subsection{Preliminaries} \label{secAlgPrelim}

Recall \cite{DiFCon97} that the zero-grade descendant fields may be characterised through their affine \opes{},
\begin{equation} \label{eqnAffPrimOPE}
\radord{\func{J}{z} \func{\phi^{\brac{m}}}{w}} = \frac{\func{\bigl( J_0 \phi^{\brac{m}} \bigr)}{w}}{z-w} + \ldots,
\end{equation}
where $\func{J}{z}$ is an (arbitrary) affine current and $\func{\bigl( J_0 \phi^{\brac{m}} \bigr)}{w}$ denotes the (zero-grade) descendant field corresponding to $J_0 \ket{\phi^{\brac{m}}}$.  We would therefore expect to obtain commutation relations of the form
\begin{equation}
\comm{J_r}{\phi^{\brac{m}}_s} = \bigl( J_0 \phi^{\brac{m}} \bigr)_{r+s}.
\end{equation}
However, it is not clear whether this expectation is correct, because we have not justified the assumption that $\func{J}{z}$ and $\func{\phi^{\brac{m}}}{w}$ commute (are mutually bosonic) in \opes{}.

Let us suppose therefore that
\begin{equation}
\radord{\func{J}{z} \func{\phi^{\brac{m}}}{w}} = \mu_{J,m} \radord{\func{\phi^{\brac{m}}}{w} \func{J}{z}},
\end{equation}
for some constants $\mu_{J,m}$.  The corresponding mode relations are then
\begin{equation}
\dcomm{J_r}{\phi^{\brac{m}}_s} \equiv J_r \phi^{\brac{m}}_s - \mu_{J,m} \phi^{\brac{m}}_s J_r = \bigl( J_0 \phi^{\brac{m}} \bigr)_{r+s}.
\end{equation}
We set $\dcomm{\phi^{\brac{m}}_s}{J_r} = - \mu_{J,m}^{-1} \dcomm{J_r}{\phi^{\brac{m}}_s}$, and consider the constraints imposed by the corresponding Jacobi identity:
\begin{equation}
\dcomm{J^a_r}{\dcomm{J^b_s}{\phi^{\brac{m}}_t}} + \mu_{J^a,m} \dcomm{J^b_s}{\dcomm{\phi^{\brac{m}}_t}{J^a_r}} + \mu_{J^a,m} \mu_{J^b,m} \dcomm{\phi^{\brac{m}}_t}{\comm{J^a_r}{J^b_s}} = 0.
\end{equation}
This becomes
\begin{equation}
\bigl( J^a_0 J^b_0 \phi^{\brac{m}} \bigr)_{r+s+t} - \bigl( J^b_0 J^a_0 \phi^{\brac{m}} \bigr)_{r+s+t} = \sum_c \frac{\mu_{J^a,m} \mu_{J^b,m}}{\mu_{J^c,m}} \strconst{J^a}{J^b}{J^c} \bigl( J^c_0 \phi^{\brac{m}} \bigr)_{r+s+t},
\end{equation}
upon expanding the affine commutator in terms of the structure constants of $\func{\alg{sl}}{2}$:
\begin{equation}
\comm{J^a_r}{J^b_s} = \sum_c \strconst{J^a}{J^b}{J^c} J^c_{r+s} + r \killing{J^a}{J^b} \delta_{r+s,0} k.  
\end{equation}
The constraints are therefore that
\begin{equation}
\mu_{J^a,m} \mu_{J^b,m} = \mu_{J^c,m} \qquad \text{if} \qquad \strconst{J^a}{J^b}{J^c} \neq 0.
\end{equation}

Up to permutation, the non-vanishing structure constants are $\strconst{E}{F}{H} = 1$ and $\strconst{H}{E}{E} = \strconst{F}{H}{F} = 2$ (\eqnref{eqnSL2Basis}), so the constraints give
\begin{equation} \label{eqnJacIdReq}
\mu_{H,m} = 1 \qquad \text{and} \qquad \mu_{E,m} = \mu_{F,m}^{-1}.
\end{equation}
In other words, the Jacobi identity requires $\func{H}{z}$ to be mutually bosonic with each $\func{\phi^{\brac{m}}}{w}$, but does \emph{not} determine the analogous behaviour of $\func{E}{z}$ and $\func{F}{z}$.  The most natural solution of these constraints is to (arbitrarily) choose $\mu_{E,m} = \mu_{F,m} = 1$, in agreement with our original expectation.  However, as we will shortly see, this choice must be consistent with another consideration, and the natural choice in that case leads to a contradiction with the choice made above.  We will therefore proceed in full generality so as to completely understand this intriguing situation.

A chiral algebra comes equipped with an involutive antilinear antiautomorphism defining the adjoint on representations.  In other words, chiral algebras are not just (graded) associative algebras, but (graded) \emph{$^*$-algebras}.  We therefore need to be able to extend the adjoint of $\func{\affine{\alg{sl}}}{2}$ to the modes $\phi^{\brac{n}}_r$ in a manner consistent with the algebraic structure of the extended theory.  The full structure of the extended theory will not be derived until \secref{secAlgGCRs}, but the part we are currently investigating places strong constraints on this extended adjoint nonetheless.

Consideration of the conformal dimension and $\func{\alg{sl}}{2}$-weight force
\begin{equation}
\bigl( \phi^{\brac{m}}_s \bigr)^{\dag} = \eps_m \phi^{\brac{k-m}}_{-s},
\end{equation}
where $\eps_m$ is a constant (which we take to be independent of $s$).  The involutive nature of the adjoint then requires that
\begin{equation} \label{eqnInvolConstraint}
\eps_m^* \eps_{k-m} = 1.
\end{equation}
The natural choice in this case is therefore to take $\eps_m = 1$ for all $m$.  However, this leads to the promised contradiction with $\mu_{E,m} = \mu_{F,m} = 1$, which is detailed in the following computation:
\begin{align}
\comm{E_r}{\phi^{\brac{m}}_s}^{\dag} &= \bigl[ m \brac{k+1-m} \bigr]^{1/2} \bigl( \phi^{\brac{m-1}}_{r+s} \bigr)^{\dag} = \bigl[ m \brac{k+1-m} \bigr]^{1/2} \phi^{\brac{k+1-m}}_{-r-s} \notag \\
&= \comm{F_{-r}}{\phi^{\brac{k-m}}_{-s}} = -\comm{\bigl( \phi^{\brac{m}}_s \bigr)^{\dag}}{E_r^{\dag}}.
\end{align}
This demonstrates that $\eps_m = 1$ does \emph{not} define an antiautomorphism of the extended theory when $\mu_{E,m} = \mu_{F,m} = 1$.

By repeating this computation (and others) for general $\eps_m$ and $\mu_{E,m}$, $\mu_{F,m}$, we find that the extended adjoint thus defined avoids this contradiction precisely when
\begin{equation}
\eps_{m-1} = - \mu_{E,k-m} \eps_m \qquad \text{and} \qquad \eps_{m+1} = - \mu_{F,k-m} \eps_m.
\end{equation}
Of course, this does not itself imply that we have constructed an antiautomorphism of the extended algebra (we will not even finish the derivation of this algebra until \secref{secAlgGCRs}), merely that the above constraints are necessary to obtain an antiautomorphism.  We note that these constraints imply that $\mu_E \equiv \mu_{E,m}$ and $\mu_F \equiv \mu_{F,m}$ are independent of $m$:
\begin{equation}
\mu_{E,k-m} = - \eps_{m-1} \eps_m^{-1} = \mu_{F,k+1-m}^{-1} = \mu_{E,k+1-m},
\end{equation}
by \eqnref{eqnJacIdReq}.  It follows that $\eps_{m} = \brac{- \mu_F}^m \eps_0$, hence that
\begin{equation}
\brac{- \mu_E}^k = \abs{\eps_0}^2 > 0,
\end{equation}
by \eqnref{eqnInvolConstraint}.

Setting $\mu_{E,m} = \mu_{F,m} = 1$ is therefore consistent with $\eps_m = \brac{-1}^m \eps_0$ if $k$ is even, but not when $k$ is odd, and never with $\eps_m = 1$.  Likewise, the choice $\eps_m = 1$ is consistent with $\mu_{E,m} = \mu_{F,m} = -1$ for all $k$, but not with $\mu_{E,m} = \mu_{F,m} = 1$.  In other words, the most convenient choices we can make to describe the algebra of $\phi$-modes are the following:
\begin{enumerate}
\item For all $k$, define $\bigl( \phi^{\brac{m}}_s \bigr)^{\dag} = \phi^{\brac{k-m}}_{-s}$ so that the $\func{\phi^{\brac{m}}}{w}$ are mutually bosonic with respect to $\func{H}{z}$ but mutually fermionic with respect to $\func{E}{z}$ and $\func{F}{z}$. The defining relations between the $H_r$ and the $\phi^{\brac{m}}_s$ are therefore commutation relations whereas those between the $E_r$ or $F_r$ and the $\phi^{\brac{m}}_s$ are anticommutation relations. \label{ChoiceFermion}
\item Define $\bigl( \phi^{\brac{m}}_s \bigr)^{\dag} = \brac{-1}^{k-m} \phi^{\brac{k-m}}_{-s}$ so that the $\func{\phi^{\brac{m}}}{w}$ are mutually bosonic with respect to all affine fields when $k$ is even, but mutually fermionic with respect to $\func{E}{z}$ and $\func{F}{z}$ when $k$ is odd.  The defining relations between the $H_r$ and the $\phi^{\brac{m}}_s$ are therefore commutation relations whereas those between the $E_r$ or $F_r$ and the $\phi^{\brac{m}}_s$ are commutation or anticommutation relations according to the parity of $k$. \label{ChoiceBoson}
\end{enumerate}
We prefer to work with choice (\ref{ChoiceFermion}) in which the adjoint and defining relations are independent of $k$.  However, in specific examples with $k$ even, it is often preferable to work with choice (\ref{ChoiceBoson}) because the formulation involving bosonic \opes{} facilitates comparison with other theories (as we will do in \secref{secExamples} for example).

When $k$ is even, the equivalence of these two choices is easy to establish with the help\footnote{We could also establish this with ``Klein factors'', but not as elegantly as in the present development.} of an auxiliary operator $\mathcal{F}$ satisfying
\begin{equation}
\mathcal{F} E_r = -E_r \mathcal{F}, \qquad \mathcal{F} H_r = H_r \mathcal{F}, \qquad \mathcal{F} F_r = -F_r \mathcal{F},\quad \text{and} \quad  \mathcal{F}^2 = \id.
\end{equation}
Distinguishing the $\phi$-modes under choice (\ref{ChoiceBoson}) by tildes, this equivalence may be explicitly described by
\begin{equation} \label{eqnFDef}
\phi^{\brac{m}}_s = \widetilde{\phi}^{\brac{m}}_s \mathcal{F} \qquad \Longleftrightarrow \qquad \widetilde{\phi}^{\brac{m}}_s = \phi^{\brac{m}}_s \mathcal{F}.
\end{equation}
Then (for instance),
\begin{align}
\sqbrac{E_r \phi^{\brac{m}}_s + \phi^{\brac{m}}_s E_r} \mathcal{F} &= \sqbrac{m \brac{k+1-m}}^{1/2} \phi^{\brac{m-1}}_{r+s} \mathcal{F} \notag \\
\Rightarrow \qquad E_r \widetilde{\phi}^{\brac{m}}_s - \widetilde{\phi}^{\brac{m}}_s E_r &= \sqbrac{m \brac{k+1-m}}^{1/2} \widetilde{\phi}^{\brac{m-1}}_{r+s}.
\end{align}
In this way, we transform anticommutation relations into commutation relations.

We also need to settle the manner in which $\mathcal{F}$ commutes with the $\phi^{\brac{m}}_s$.  Supposing that $\mathcal{F} \phi^{\brac{m}}_s = \omega_m \phi^{\brac{m}}_s \mathcal{F}$, we have
\begin{align}
\mathcal{F} \sqbrac{E_r \phi^{\brac{m}}_s + \phi^{\brac{m}}_s E_r} &= \sqbrac{m \brac{k+1-m}}^{1/2} \mathcal{F} \phi^{\brac{m-1}}_{r+s} \notag\\
\Rightarrow \qquad - \omega_m \sqbrac{E_r \widetilde{\phi}^{\brac{m}}_s - \widetilde{\phi}^{\brac{m}}_s E_r} &= \omega_{m-1} \sqbrac{m \brac{k+1-m}}^{1/2} \widetilde{\phi}^{\brac{m-1}}_{r+s}.
\end{align}
Therefore $\omega_m = - \omega_{m-1}$.  Considering the corresponding relations for the other affine modes yields no further constraints, hence we are free to choose $\omega_m = \brac{-1}^m$.  In other words, $\mathcal{F}$ and $\phi^{\brac{m}}_s$ may be taken to commute when $m$ is even, but anticommute when $m$ is odd.

Finally, we may take $\mathcal{F}$ to be self-adjoint.  This follows from
\begin{equation}
\mathcal{F}^{\dag} \phi^{\brac{m}}_s = \bigl( \phi^{\brac{k-m}}_{-s} \mathcal{F} \bigr)^{\dag} = \bigl( \widetilde{\phi}^{\brac{k-m}}_{-s} \bigr)^{\dag} = \brac{-1}^m \widetilde{\phi}^{\brac{m}}_s = \brac{-1}^m \phi^{\brac{m}}_s \mathcal{F} = \mathcal{F} \phi^{\brac{m}}_s,
\end{equation}
remembering that $k$ is assumed to be even.

To summarise then, we have an extension of the chiral algebra $\uealg{\func{\affine{\alg{sl}}}{2}_k}$ by the modes $\phi^{\brac{m}}_s$ of the simple current.  We choose to extend the adjoint antiautomorphically as in choice (\ref{ChoiceFermion}),
\begin{equation} \label{eqnAdjoint}
\bigl( \phi^{\brac{m}}_s \bigr)^{\dag} = \phi^{\brac{k-m}}_{-s},
\end{equation}
which determines the commutativity of the \opes{} to be
\begin{equation} \label{eqnAffPrimOPEComm}
\begin{split}
\radord{\func{E}{z} \func{\phi^{\brac{m}}}{w}} &= - \radord{\func{\phi^{\brac{m}}}{w} \func{E}{z}}, \\
\radord{\func{H}{z} \func{\phi^{\brac{m}}}{w}} &= \phantom{-} \radord{\func{\phi^{\brac{m}}}{w} \func{H}{z}}, \\
\text{and} \qquad \radord{\func{F}{z} \func{\phi^{\brac{m}}}{w}} &= - \radord{\func{\phi^{\brac{m}}}{w} \func{F}{z}}.
\end{split}
\end{equation}
The corresponding mode relations are therefore
\begin{equation} \label{eqnAffPrimComm}
\begin{split}
\acomm{E_r}{\phi^{\brac{m}}_s} &= \bigl[ m \brac{k+1-m} \bigr]^{1/2} \phi^{\brac{m-1}}_{r+s}, \\
\comm{H_r}{\phi^{\brac{m}}_s} &= \brac{k-2m} \phi^{\brac{m}}_{r+s}, \\
\text{and} \qquad \acomm{F_r}{\phi^{\brac{m}}_s} &= \bigl[ \brac{m+1} \brac{k-m} \bigr]^{1/2} \phi^{\brac{m+1}}_{r+s},
\end{split}
\end{equation}
where $\acomm{\cdot}{\cdot}$ denotes an anticommutator.  The equivalence of this extension with that which would be obtained through choice (\ref{ChoiceBoson}) (when $k$ is even) is effected by introducing a self-adjoint operator $\mathcal{F}$, as in \eqnref{eqnFDef}, which squares to $\id$, commutes with the $H_r$ and the $\phi^{\brac{m}}_s$ with $m$ even, and anticommutes with the $E_r$, $F_r$, and the $\phi^{\brac{m}}_s$ with $m$ odd.

To complete the description of this extended chiral algebra, we have to determine the algebraic relations between the modes $\phi^{\brac{m}}_r$ and $\phi^{\brac{n}}_s$.  These relations are derived from the corresponding \opes{} whose computation we now turn to.

\subsection{Operator Product Expansions} \label{secAlgOPEs}

\begin{figure}
\begin{center}
\includegraphics[width=10cm]{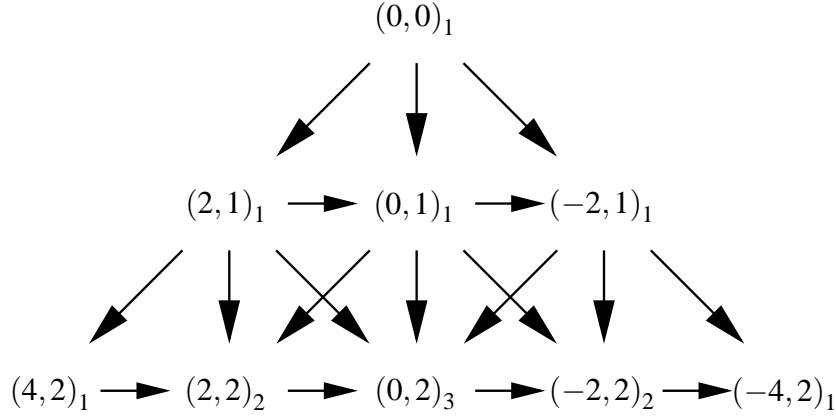}
\caption{The first few grades of a generic vacuum $\func{\affine{\alg{sl}}}{2}$-module (this triangular pattern continues until we arrive at the affine singular vectors at grade $k+1$).  The states are written in the form $\pair{\mu}{h}$ where $\mu$ denotes the $\func{\alg{sl}}{2}$-weight, $h$ denotes the conformal dimension, and the subscript denotes the multiplicity.} \label{figSU2VacModule}
\end{center}
\end{figure}

Let us consider the \opes{}
\begin{equation} \label{eqnGeneralOPE}
\radord{\func{\phi^{\brac{m}}}{z} \func{\phi^{\brac{n}}}{w}} = \sum_{j=0}^{\infty} \func{A^{\brac{m,n ; j}}}{w} \brac{z-w}^{j - k/2},
\end{equation}
for $m, n = 0, 1, \ldots, k$, where we have chosen $\func{A^{\brac{m,n ; j}}}{w}$ to have conformal dimension $j$.  Since $\phi \times \phi = \psi_0$, the fields $\func{A^{\brac{m,n ; j}}}{w}$ may be expressed in terms of fields descended from the identity field, and we will compute the form of these expressions by considering the corresponding states:
\begin{equation} \label{eqnAStates}
\ket{A^{\brac{m,n ; j}}} = \lim_{w \rightarrow 0} \oint_w \radord{\func{\phi^{\brac{m}}}{z} \func{\phi^{\brac{n}}}{w}} \brac{z-w}^{k/2 - j - 1} \frac{\dd z}{2 \pi \ii} \ket{0} = \phi^{\brac{m}}_{k/4 - j} \phi^{\brac{n}}_{-k/4} \ket{0}.
\end{equation}
To do this, we require some structural knowledge of the vacuum $\func{\affine{\alg{sl}}}{2}$-module.  We indicate this structure in \figref{figSU2VacModule} for the first few grades.  The three states at grade $1$ will be denoted by $\ket{E}$, $\ket{H}$ and $\ket{F}$, as they are the states corresponding to the three affine current fields.

\psfrag{A}[][]{$\pair{0}{0}_1$}
\psfrag{B}[][]{$\pair{2}{1}_1$}
\psfrag{C}[][]{$\pair{0}{1}_1$}
\psfrag{D}[][]{$\pair{-2}{1}_1$}
\psfrag{E}[][]{$\pair{4}{2}_1$}
\psfrag{F}[][]{$\pair{2}{2}_2$}
\psfrag{G}[][]{$\pair{0}{2}_3$}
\psfrag{H}[][]{$\pair{-2}{2}_2$}
\psfrag{J}[][]{$\pair{-4}{2}_1$}

Taking $j = 0$ in \eqnref{eqnAStates}, we see that $\ket{A^{\brac{m,n ; 0}}}$ must be a state in the vacuum $\func{\affine{\alg{sl}}}{2}$-module of conformal dimension $0$ and $\func{\alg{sl}}{2}$-weight $2 \brac{k-m-n}$.  It is therefore proportional to $\delta_{m+n , k} \ket{0}$, and the proportionality constant may be evaluated by comparing
\begin{equation}
\braket{0}{A^{\brac{m,n ; 0}}} = \bracket{0}{\phi^{\brac{m}}_{k/4} \phi^{\brac{n}}_{-k/4}}{0} = \braket{\phi^{\brac{k-m}}}{\phi^{\brac{n}}} = \delta_{m+n , k}
\end{equation}
and $\braket{0}{0} = 1$.  It follows that $\ket{A^{\brac{m,n ; 0}}} = \delta_{m+n , k} \ket{0}$.

If $m+n = k-1$, then $\ket{A^{\brac{m,n ; 1}}}$ is a state of conformal dimension $1$ and $\func{\alg{sl}}{2}$-weight $2$, hence is proportional to $\ket{E}$.  Again, we evaluate the constant of proportionality by comparing
\begin{equation}
\braket{E}{E} = \bracket{0}{F_1 E_{-1}}{0} = \bracket{0}{-H_0 + k}{0} = k,
\end{equation}
where we have used \eqnref{eqnAffCommRels}, and
\begin{align}
\braket{E}{A^{\brac{m,n ; 1}}} &= \bracket{0}{F_1 \phi^{\brac{m}}_{k/4 - 1}}{\phi^{\brac{n}}} = \bracket{0}{\bigl[ \brac{m+1} \brac{k-m} \bigr]^{1/2} \phi^{\brac{m+1}}_{k/4}}{\phi^{\brac{n}}} \notag \\
&= \bigl[ \brac{m+1} \brac{n+1} \bigr]^{1/2} \delta_{k-m-1,n},
\end{align}
where we have used \eqnDref{eqnSU2Normalisation}{eqnAffPrimComm}.  It follows that
\begin{align}
\ket{A^{\brac{m,n ; 1}}} &= \frac{1}{k} \bigl[ \brac{m+1} \brac{n+1} \bigr]^{1/2} \ket{E} & &\text{if $m+n = k-1$.}
\intertext{Similar calculations give}
\ket{A^{\brac{m,n ; 1}}} &= \frac{1}{2k} \brac{n-m} \ket{H} & &\text{if $m+n = k$,} \\
\text{and} \qquad \ket{A^{\brac{m,n ; 1}}} &= \frac{1}{k} \bigl[ mn \bigr]^{1/2} \ket{F} & &\text{if $m+n = k+1$.}
\end{align}
Clearly $\ket{A^{\brac{m,n ; 1}}}$ vanishes otherwise.

We can repeat these computations to determine $\ket{A^{\brac{m,n ; 2}}}$.  However, we note that the relevant states in \figref{figSU2VacModule} have non-trivial multiplicities, and some of them will be singular when $k=1$.  It is therefore appropriate to choose bases of the corresponding state spaces accordingly.  We will use the following bases for the spaces of weights $2$, $0$ and $-2$:
\begin{equation}
\Bigl\{ E_{-2} \ket{0} , \Delta^+_{-2} \ket{0} \Bigr\}, \quad \Bigl\{ L_{-2} \ket{0} , H_{-2} \ket{0} , \Delta^0_{-2} \ket{0} \Bigr\} \quad \text{and} \quad \Bigl\{ F_{-2} \ket{0} , \Delta^-_{-2} \ket{0} \Bigr\},
\end{equation}
where
\begin{equation}
\setlength{\extrarowheight}{1mm}
\begin{array}{c}
\Delta^+_{-2} = H_{-1} E_{-1} + E_{-1} H_{-1}, \qquad \Delta^-_{-2} = H_{-1} F_{-1} + F_{-1} H_{-1} \\
\text{and} \qquad \Delta^0_{-2} = H_{-1} H_{-1} - E_{-1} F_{-1} - F_{-1} E_{-1}.
\end{array}
\end{equation}
It is not hard to check that each of these bases consists of mutually orthogonal states, and that they have the nice property that the singular vector is a basis element (the $\Delta^{\bullet}_{-2} \ket{0}$) when $k=1$, hence can be easily removed.

We therefore compute the $\ket{A^{\pair{m}{n}}_2}$ as before, omitting the details.  The non-zero results are as follows:
\begin{align}
m+n = k-2 &: & \ket{A^{\brac{m,n ; 2}}} &= \frac{1}{2k \brac{k-1}} \bigl[ \brac{m+1} \brac{m+2} \brac{n+1} \brac{n+2} \bigr]^{1/2} E_{-1} E_{-1} \ket{0}, \notag \\
m+n = k-1 &: & \ket{A^{\brac{m,n ; 2}}} &= \frac{1}{k} \bigl[ \brac{m+1} \brac{n+1} \bigr]^{1/2} \sqbrac{\frac{1}{2} E_{-2} + \frac{n-m}{4 \brac{k-1}} \Delta^+_{-2}} \ket{0}, \notag \\
m+n = k &: & \ket{A^{\brac{m,n ; 2}}} &= \sqbrac{\frac{k+2}{6} L_{-2} + \frac{n-m}{4k} H_{-2} + \brac{\frac{1}{12} - \frac{mn}{2k \brac{k-1}}} \Delta^0_{-2}} \ket{0}, \\
m+n = k+1 &: & \ket{A^{\brac{m,n ; 2}}} &= \frac{1}{k} \bigl[ mn \bigr]^{1/2} \sqbrac{\frac{1}{2} F_{-2} + \frac{n-m}{4 \brac{k-1}} \Delta^-_{-2}} \ket{0}, \notag \\
m+n = k+2 &: & \ket{A^{\brac{m,n ; 2}}} &= \frac{1}{2k \brac{k-1}} \bigl[ m \brac{m-1} n \brac{n-1} \bigr]^{1/2} F_{-1} F_{-1} \notag \ket{0}.
\end{align}
When $k=1$, these results are modified by simply dropping the modes whose coefficients diverge (as these modes correspond to singular vectors).

This can obviously be extended to $\ket{A^{\brac{m,n ; j}}}$ for $j > 2$, though with significant increase in effort as $j$ increases.  We would like to point out one important (and useful) feature of this extension:  For arbitrary $m$ and $n$, the first $\ket{A^{\brac{m,n ; j}}}$ which does not identically vanish is that with $j = \abs{k-m-n}$.  This may be seen by noting that the vacuum descendant of $\func{\alg{sl}}{2}$-weight $2 \brac{k-m-n}$ and minimal conformal dimension is obtained by repeatedly acting on the vacuum with either $E_{-1}$ or $F_{-1}$.  This first state is therefore proportional to either $E_{-1}^j \ket{0}$ or $F_{-1}^j \ket{0}$ (depending on whether the weight is positive or negative).  This proportionality constant is easily determined (as above), giving
\begin{equation} \label{eqnPropConsts}
\ket{A^{\brac{m,n ; \abs{k-m-n}}}} = 
\begin{cases}
\displaystyle \frac{\brac{m+n} !}{k !} \sqbrac{\binom{k-n}{m} \binom{k-m}{n}}^{1/2} E_{-1}^{k-m-n} \ket{0} & \text{if $m+n \leqslant k$,} \\
\displaystyle \frac{\brac{2k-m-n} !}{k !} \sqbrac{\binom{m}{k-n} \binom{n}{k-m}}^{1/2} F_{-1}^{m+n-k} \ket{0} & \text{if $m+n \geqslant k$.}
\end{cases}
\end{equation}
Note that these constants are invariant under $\pair{m}{n} \rightarrow \pair{n}{m}$ and $\pair{m}{n} \rightarrow \pair{k-m}{k-n}$.

It is tempting to conclude from the above results that (for example)
\begin{equation}
\func{A^{\brac{m,n ; 2}}}{w} = \frac{1}{2k \brac{k-1}} \bigl[ m \brac{m-1} n \brac{n-1} \bigr]^{1/2} \normord{\func{F}{w} \func{F}{w}} \qquad \text{if $m+n = k+2$.}
\end{equation}
However, such conclusions are not yet justified, because we have no \emph{a priori} guarantee that these conclusions will preserve the associativity of the \opes{}.  Associativity is a fundamental requirement of a consistent conformal field theory, but it need not be automatically satisfied.  As mentioned in \secref{secOrg} (see also \cite{RidMin06}), it is sometimes necessary (and often convenient) to introduce an operator $\mathcal{S}$ into the \ope{} of two fields in order to ensure the associativity of the operator product algebra.  This will not be in conflict with the state-field correspondence if this $\mathcal{S}$-operator acts as (a multiple of) the identity on the vacuum module.  We therefore turn to a careful (and rigorous) study of associativity.

\subsection{Associativity} \label{secAlgAssoc}

We begin with the commutativity of the radially-ordered product.  As noted above, the first contributing term in the \ope{} (\ref{eqnGeneralOPE}) corresponds to $j = \abs{k-m-n}$.  It follows that the most general form of commutativity may be expressed as
\begin{equation} \label{eqnROEComm}
\brac{z-w}^{k/2 - \abs{k-m-n}} \radord{\func{\phi^{\brac{m}}}{z} \func{\phi^{\brac{n}}}{w}} = \mu_{m,n} \radord{\func{\phi^{\brac{n}}}{w} \func{\phi^{\brac{m}}}{z}} \brac{w-z}^{k/2 - \abs{k-m-n}},
\end{equation}
where $\mu_{m,n} = \mu_{n,m}^{-1}$ is a phase (compare \eqnref{eqnAffPrimOPEComm} for an example exhibiting non-trivial such phases).  We do not cancel the powers of $z-w$ and $w-z$ above as the resulting power of $-1$ will not be integral when $k$ is odd.

Substituting \eqnref{eqnGeneralOPE} into \eqnref{eqnROEComm}, we derive the relation
\begin{equation}
\func{A^{\brac{m,n ; j}}}{w} = \mu_{m,n} \brac{-1}^{k-m-n} \sum_{i=0}^j \frac{\brac{-1}^{j-i}}{i !} \partial^i \func{A^{\brac{n,m ; j-i}}}{w}.
\end{equation}
Putting $j = \abs{k-m-n}$, only one term ($i = 0$) contributes on the right-hand-side, and the corresponding relation between states becomes
\begin{equation}
\ket{A^{\brac{m,n ; \abs{k-m-n}}}} = \mu_{m,n} \ket{A^{\brac{n,m ; \abs{k-m-n}}}}.
\end{equation}
It now follows from \eqnref{eqnPropConsts} that the two states appearing in this equation are identical and non-vanishing, hence $\mu_{m,n} = 1$ for all $m$ and $n$.  We therefore drop these phases in what follows.

We now consider the consistency of the operator product algebra.  In particular, we investigate its associativity by analysing what happens when we use \eqnref{eqnROEComm} twice to obtain
\begin{multline}
\oint_z \radord{\func{\phi^{\brac{m}}}{x} \func{\phi^{\brac{n}}}{z} \func{\phi^{\brac{p}}}{w}} \brac{x-z}^{k/2 - \gamma} \brac{x-w}^{k/2 - \abs{k-m-p}} \brac{z-w}^{k/2 - \abs{k-n-p}} \frac{\dd x}{2 \pi \ii} \\*
= \oint_z \radord{\func{\phi^{\brac{p}}}{w} \func{\phi^{\brac{m}}}{x} \func{\phi^{\brac{n}}}{z}} \brac{x-z}^{k/2 - \gamma} \brac{w-x}^{k/2 - \abs{k-m-p}} \brac{w-z}^{k/2 - \abs{k-n-p}} \frac{\dd x}{2 \pi \ii},
\end{multline}
where $\gamma \in \ZZ$ is arbitrary.  Inserting our \ope{} (\ref{eqnGeneralOPE}) and evaluating the contour integral gives
\begin{multline}
\sum_{j=0}^{\gamma - 1} \binom{k/2 - \abs{k-m-p}}{\gamma - j - 1} \Biggl[ \radord{\func{A^{\brac{m,n ; j}}}{z} \func{\phi^{\brac{p}}}{w}} \brac{z-w}^{k - \abs{k-m-p} - \abs{k-n-p} + j - \gamma + 1} \Biggr. \\*
\Biggl. - \brac{-1}^{\gamma - j - 1} \radord{ \func{\phi^{\brac{p}}}{w} \func{A^{\brac{m,n ; j}}}{z}} \brac{w-z}^{k - \abs{k-m-p} - \abs{k-n-p} + j - \gamma + 1} \Biggr] = 0.
\end{multline}
Note that the common exponent of $z-w$ and $w-z$ is an integer, so we may factor out this power of $z-w$, leaving the second term with a (well-defined) relative sign:  $\brac{-1}^{k - \abs{k-m-p} - \abs{k-n-p}} = \brac{-1}^{k-m-n}$.  Note also that this equation now has the form of a matrix equation $M . v = 0$, where $M$ is lower-triangular (its entries are the binomial coefficients appearing above) with ones on the diagonal, hence invertible.  We therefore conclude that associativity requires that
\begin{equation} \label{eqnAPrimOPEComm}
\radord{\func{A^{\brac{m,n ; j}}}{z} \func{\phi^{\brac{p}}}{w}} = \brac{-1}^{k-m-n} \radord{ \func{\phi^{\brac{p}}}{w} \func{A^{\brac{m,n ; j}}}{z}}.
\end{equation}

The field $\func{A^{\brac{m,n ; j}}}{z}$ has $\func{\alg{sl}}{2}$-weight $2 \brac{k-m-n}$, so it must be decomposable into terms containing an even or an odd number of $\func{E}{z}$ and $\func{F}{z}$ (and their derivatives), according as to whether $k-m-n$ is even or odd (respectively).  Comparing \eqnref{eqnAPrimOPEComm} with \eqnref{eqnAffPrimOPEComm} now, we see that this is exactly the commutativity that we would expect, hence it is not necessary to introduce $\mathcal{S}$-operators to preserve associativity.  Nevertheless, it turns out to be convenient to introduce operators $\mathcal{S}_{m,n}$ into our simple current \opes{} (we will see why shortly).  The lack of necessity for introducing these operators translates into the fact that they must commute with every mode of the extended symmetry algebra (to be defined next), hence will act as multiples of the identity on the Verma modules of the extended algebra (to be introduced in \secref{secRepHWMs}).

\subsection{Generalised Commutation Relations} \label{secAlgGCRs}

Having settled the question of associativity, it is now convenient to \emph{redefine} the \ope{} (\ref{eqnGeneralOPE}) to have the form
\begin{equation} \label{eqnOPEs}
\radord{\func{\phi^{\brac{m}}}{z} \func{\phi^{\brac{n}}}{w}} = \mathcal{S}_{m,n} \sum_{j=0}^{\infty} \func{A^{\brac{m,n ; j}}}{w} \brac{z-w}^{j - k/2} \equiv \frac{\mathcal{S}_{m,n}}{\brac{z-w}^{k/2}} \func{\Xi_{m,n}}{z,w}.
\end{equation}
In other words, we explicitly factor out the $\mathcal{S}_{m,n}$ from the definition of the fields $\func{A^{\brac{m,n ; j}}}{w}$.  The eigenvalues of the $\mathcal{S}_{m,n}$ on the vacuum are given by \eqnref{eqnPropConsts} as
\begin{equation} \label{eqnSValues}
\mathcal{S}_{m,n} \ket{0} = 
\begin{cases}
\displaystyle \frac{\brac{m+n} !}{k !} \sqbrac{\binom{k-n}{m} \binom{k-m}{n}}^{1/2} \ket{0} & \text{if $m+n \leqslant k$,} \\
\displaystyle \frac{\brac{2k-m-n} !}{k !} \sqbrac{\binom{m}{k-n} \binom{n}{k-m}}^{1/2} \ket{0} & \text{if $m+n \geqslant k$,}
\end{cases}
\end{equation}
and the $\func{\Xi_{m,n}}{z,w}$ which do not vanish to second order are (compare \secref{secAlgOPEs}):
\begin{equation}
\begin{split}
m+n=k-2 &: \quad \normord{\func{E}{w} \func{E}{w}} \brac{z-w}^2 + \ldots \\
m+n=k-1 &: \quad \func{E}{w} \brac{z-w} + \sqbrac{\frac{1}{2} \func{\partial E}{w} + \frac{n-m}{4 \brac{k-1}} \func{\Delta^+}{w}} \brac{z-w}^2 + \ldots \\
m+n=k &: \quad 1 + \frac{n-m}{2k} \func{H}{w} \brac{z-w} \\*
& \mspace{0mu} + \sqbrac{\frac{k+2}{6} \func{T}{w} + \frac{n-m}{4k} \func{\partial H}{w} + \brac{\frac{1}{12} - \frac{mn}{2k \brac{k-1}}} \func{\Delta^0}{w}} \brac{z-w}^2 + \ldots \\
m+n=k+1 &: \quad \func{F}{w} \brac{z-w} + \sqbrac{\frac{1}{2} \func{\partial F}{w} + \frac{n-m}{4 \brac{k-1}} \func{\Delta^-}{w}} \brac{z-w}^2 + \ldots \\
m+n=k+2 &: \quad \normord{\func{F}{w} \func{F}{w}} \brac{z-w}^2 + \ldots
\end{split}
\end{equation}
When $k=1$, these series are modified by simply ignoring all fields whose coefficients contain the divergent factor $\brac{k-1}^{-1}$.  We note that \eqnref{eqnPropConsts} implies that the leading term of $\func{\Xi_{m,n}}{z,w}$ is 
\begin{equation} \label{eqnLeadingTerm}
\normord{\overbrace{\func{E}{w} \cdots \func{E}{w}}^{k-m-n}} \brac{z-w}^{k-m-n} \qquad \text{or} \qquad \normord{\overbrace{\func{F}{w} \cdots \func{F}{w}}^{m+n-k}} \brac{z-w}^{m+n-k},
\end{equation}
according as to whether $m+n \leqslant k$ or $m+n \geqslant k$ (respectively).

The \gcrs{} are obtained from these \opes{} by evaluating
\begin{equation} \label{eqnDefRGamma}
\func{R^{\brac{m,n}}_{r,s}}{\gamma} = \oint_0 \oint_w \radord{\func{\phi^{\brac{m}}}{z} \func{\phi^{\brac{n}}}{w}} z^{r - k/4 + \gamma - 1} w^{s + k/4 - 1} \brac{z-w}^{k/2 - \gamma} \frac{\dd z}{2 \pi \ii} \frac{\dd w}{2 \pi \ii},
\end{equation}
where $\gamma \in \ZZ$ is arbitrary, in two distinct fashions:  Expanding as a difference of two contours,
\begin{equation}
\oint_0 \oint_w \cdots \frac{\dd z}{2 \pi \ii} \frac{\dd w}{2 \pi \ii} = \underset{\abs{z} > \abs{w}}{\oint_0 \oint_0} \cdots \frac{\dd z}{2 \pi \ii} \frac{\dd w}{2 \pi \ii} - \underset{\abs{z} < \abs{w}}{\oint_0 \oint_0} \cdots \frac{\dd z}{2 \pi \ii} \frac{\dd w}{2 \pi \ii},
\end{equation}
and using \eqnref{eqnROEComm} (with $\mu_{m,n} = 1$), or expanding directly using \eqnref{eqnOPEs}.  This procedure yields the set of generalised commutation relations (parametrised by $\gamma$, $m$, $n$, $r$ and $s$):
\begin{multline} \label{eqnGeneralGCRs}
\sum_{\ell = 0}^{\infty} \binom{\ell - k/2 + \gamma - 1}{\ell} \sqbrac{\phi^{\brac{m}}_{r - \ell} \phi^{\brac{n}}_{s + \ell} - \brac{-1}^{k-m-n + \gamma} \phi^{\brac{n}}_{s + k/2 - \gamma - \ell} \phi^{\brac{m}}_{r - k/2 + \gamma + \ell}} \\*
= \mathcal{S}_{m,n} \sum_{j=0}^{\gamma - 1} \binom{r - k/4 + \gamma - 1}{\gamma - 1 - j} A^{\brac{m,n \, ; j}}_{r+s}.
\end{multline}
Note that the modes $A^{\brac{m,n \, ; j}}_{r+s}$ refer to the fields appearing in \eqnref{eqnOPEs} (with the $\mathcal{S}_{m,n}$ explicitly factored out), not those of \eqnref{eqnGeneralOPE}.

We observe that if $k$ is even, we may choose $\gamma = k/2$, obtaining a genuine commutation or anticommutation relation (according as to the parity of $\gamma - m - n$).  For $k$ odd however, every \gcr{} will have infinitely many terms on the left hand side.  We also observe that for $\gamma \leqslant 0$, the right hand side of the \gcrs{} clearly vanish. When positive, $\gamma$ is a measure of the depth at which the \ope{} (\ref{eqnOPEs}) is probed. Indeed, $\gamma$ represents the (maximal) number of terms (singular or regular) of the \ope{} which contribute to the \gcr{}.  As these relations will be frequently used in what follows, we will find it convenient to employ a shorthand notation ``$\gcreq{\gamma}$'' to indicate that the \gcr{} of order $\gamma$ has been used to obtain an equality.

In any case, these \gcrs{} define the algebraic structure of the extended chiral algebra.  To be precise, the extended chiral algebra is the $^*$-algebra $\alg{A}_k$ generated by the modes $\phi^{\brac{m}}_r$, subject to these \gcrs{}, and equipped with the adjoint chosen in \secref{secAlgPrelim}.  The $\mathcal{S}_{m,n}$ with $m+n = k$ are generated by the $\gamma = 1$ \gcrs{}
\begin{equation}
\mathcal{S}_{m,k-m} \gcreq{1} \sum_{\ell = 0}^{\infty} \binom{\ell - k/2}{\ell} \sqbrac{\phi^{\brac{m}}_{r - \ell} \phi^{\brac{k-m}}_{\ell - r} + \phi^{\brac{k-m}}_{k/2 - r - \ell - 1} \phi^{\brac{m}}_{r - k/2 + \ell + 1}},
\end{equation}
and the other $\mathcal{S}_{m,n}$ are multiples of these (as we shall see in \secref{secSEigs}).  The affine modes are then generated by the $\gamma = 2$ \gcrs{}, for example
\begin{equation}
\mathcal{S}_{0,k-1} E_{r+s} \gcreq{2} \sum_{\ell = 0}^{\infty} \binom{\ell - k/2 + 1}{\ell} \sqbrac{\phi^{\brac{0}}_{r - \ell} \phi^{\brac{k-1}}_{s + \ell} + \phi^{\brac{k-1}}_{s + k/2 - \ell - 2} \phi^{\brac{0}}_{r - k/2 + \ell + 2}},
\end{equation}
and the affine modes commute or anticommute with the $\phi^{\brac{m}}_r$ as in \eqnref{eqnAffPrimComm}.  It follows that $\uealg{\func{\affine{\alg{sl}}}{2}_k}$ is a subalgebra of $\alg{A}_k$, just as $\uealg{\alg{Vir}_c}$ is identified as a subalgebra of $\uealg{\func{\affine{\alg{sl}}}{2}_k}$ through the Sugawara construction (\eqnref{eqnSugawara}).

We conclude by checking that the adjoint defined in \secref{secAlgPrelim} is indeed an involutive antiautomorphism of $\alg{A}_k$, completing the check of the self-consistency of the extended chiral theory.  Indeed, we find that the function $\func{R^{\brac{m,n}}_{r,s}}{\gamma}$ defining the \gcrs{} is sent to $\func{R^{\brac{k-n,k-m}}_{-s,-r}}{\gamma}$ by the adjoint (this is easiest to see with the left hand side of \eqnref{eqnGeneralGCRs}).  Hence, the adjoint preserves the set of \gcrs{} and therefore is consistently defined on $\alg{A}_k$.  We remark that this implies that $\mathcal{S}_{m,n} = \mathcal{S}_{k-n , k-m}$, which we have already noted from \eqnref{eqnSValues}.

\section{Representation Theory} \label{secRepTheory}

\subsection{Monodromy Charge} \label{secRepCharges}

We now turn to a study of the expansion of the fields $\func{\phi^{\brac{m}}}{z}$ into modes.  As is usual with non-bosonic fields, this expansion depends upon which state the field acts.  On each zero-grade descendant state $\ket{\psi_{\lambda}^{\brac{n}}}$, we therefore define a set of $k+1$ $\func{\alg{u}}{1}$-valued charges $\theta_{\lambda}^{\pair{m}{n}}$ ($m = 0, 1, \ldots, k$) as follows:  $\theta_{\lambda}^{\pair{m}{n}}$ is the negative of the leading (smallest) power of $z - w$ in $\radord{\func{\phi^{\brac{m}}}{z} \func{\psi_{\lambda}^{\brac{n}}}{w}}$ (whose coefficient is non-zero):
\begin{equation} \label{eqnOPECharge}
\radord{\func{\phi^{\brac{m}}}{z} \func{\psi_{\lambda}^{\brac{n}}}{w}} = \frac{\func{\zeta}{w}}{\brac{z-w}^{\theta_{\lambda}^{\pair{m}{n}}}} + \ldots.
\end{equation}
From the fusion rules, it might appear that this power is
\begin{equation}
\theta_{\lambda} \equiv h_k + h_{\lambda} - h_{k - \lambda} = \frac{\lambda}{2},
\end{equation}
by \eqnref{eqnHWSConfDim}, but this need not be the case in general, as evidenced by the case $\lambda = k$ which we studied in \secref{secAlgOPEs}.  Indeed, the $\func{\alg{sl}}{2}$-weight is conserved in \opes{}, so the leading term in the \ope{} (\ref{eqnOPECharge}) should involve a zero-grade descendant field of $\func{\psi_{k - \lambda}}{w}$, whose weight is $k - 2m + \lambda - 2n$.  However, such a field need not exist.

\begin{figure}
\begin{center}
\includegraphics[width=15cm]{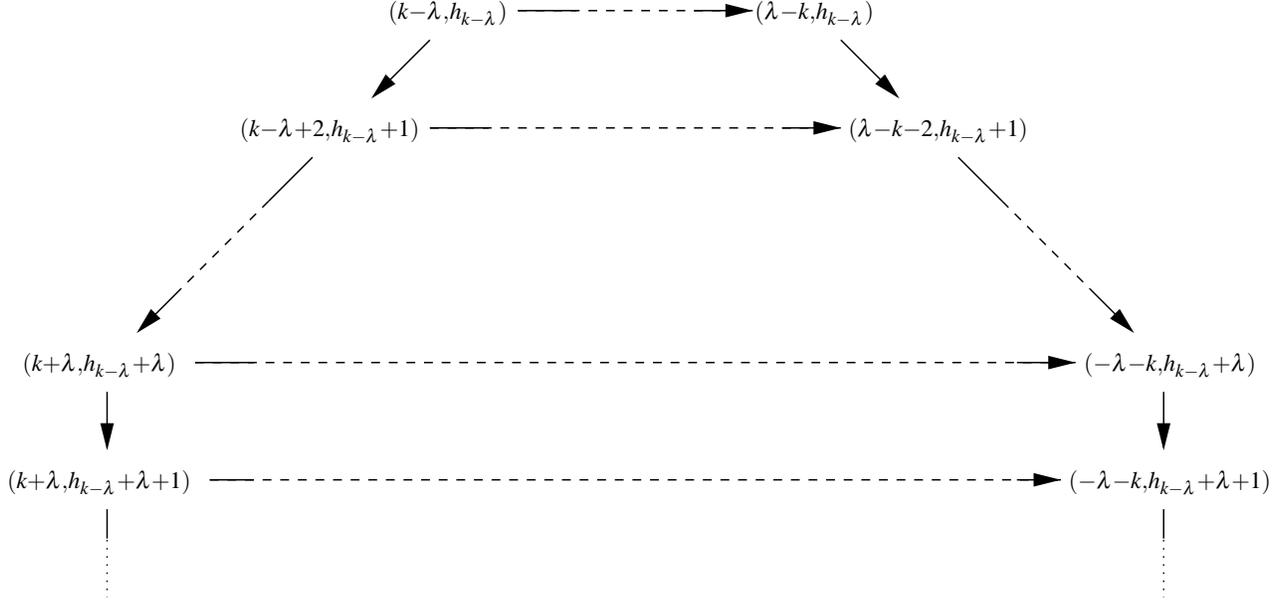}
\caption{A schematic diagram of the level $k$ irreducible $\func{\affine{\alg{sl}}}{2}$-module of highest weight $k - \lambda$.  The states are represented in the form $\pair{\mu}{h}$ where $\mu$ is the $\func{\alg{sl}}{2}$-weight and $h$ is the conformal dimension (and we have ignored any multiplicity label).} \label{figSU2Module}
\end{center}
\end{figure}

To analyse this question of existence, we consider the structure of the irreducible $\func{\affine{\alg{sl}}}{2}$-module of highest weight $k - \lambda$.  This is illustrated schematically in \figref{figSU2Module}.  We see that there will be a zero-grade descendant field of weight $k + \lambda - 2 \brac{m + n}$ if and only if $\lambda \leqslant m+n \leqslant k$.  In all other cases, the leading term in the expansion (\ref{eqnOPECharge}) will involve a descendant field at non-zero grade.  However, as $0 \leqslant m+n \leqslant k + \lambda$, these cases are easily analysed (in particular, it is not necessary to consider the affine singular vectors), and we find that
\begin{equation} \label{eqnu1Charge}
\theta_{\lambda}^{\pair{m}{n}} = 
\begin{cases}
\theta_{\lambda} - \lambda + m + n & \text{if $0 \leqslant m+n \leqslant \lambda$,} \\
\theta_{\lambda} & \text{if $\lambda \leqslant m+n \leqslant k$,} \\
\theta_{\lambda} +k - m - n & \text{if $k \leqslant m+n \leqslant k + \lambda$,}
\end{cases}
\end{equation}
where $\theta_{\lambda} = \lambda / 2$.  Note that $\theta_k^{\pair{m}{n}} = k/2 - \abs{k-m-n}$ in agreement with the general \opes{} given in \secref{secAlgGCRs}, especially \eqnref{eqnLeadingTerm}.  Of course, $\theta_{\lambda}^{\pair{m}{n}} = \theta_{\lambda} \pmod{1}$ for all $m$ and $n$ (this is a direct consequence of $\phi$ being a simple current), so the corresponding $\func{\group{U}}{1}$-valued charges are independent of $m$ (and $n$).  We shall refer to this common charge modulo $1$ as the \emph{monodromy charge}, because its negative represents the common monodromy of the $z-w$ factors in $\radord{\func{\phi^{\brac{m}}}{z} \func{\psi_{\lambda}^{\brac{n}}}{w}}$.  We note that the monodromy charge of a zero-grade descendant state is always in $\ZZ / 2$.

\psfrag{(P)}[][]{$\scriptstyle \pair{k - \lambda}{h_{k - \lambda}}$}
\psfrag{(DP)}[][]{$\scriptstyle \pair{k - \lambda + 2}{h_{k - \lambda} + 1}$}
\psfrag{(D2P)}[][]{$\scriptstyle \pair{k + \lambda}{h_{k - \lambda} + \lambda}$}
\psfrag{(D3P)}[][]{$\scriptstyle \pair{k + \lambda}{h_{k - \lambda} + \lambda + 1}$}
\psfrag{(ZG)}[][]{$\scriptstyle \pair{\lambda - k}{h_{k - \lambda}}$}
\psfrag{(DZG)}[][]{$\scriptstyle \pair{\lambda - k - 2}{h_{k - \lambda} + 1}$}
\psfrag{(D2ZG)}[][]{$\scriptstyle \pair{-\lambda - k}{h_{k - \lambda} + \lambda}$}
\psfrag{(D3ZG)}[][]{$\scriptstyle \pair{-\lambda - k}{h_{k - \lambda} + \lambda + 1}$}

The importance of assigning a monodromy charge to states is that it tells us how to expand the fields $\func{\phi^{\brac{m}}}{z}$ into modes.  To be more precise, when expanding
\begin{equation}
\func{\phi^{\brac{m}}}{z} \ket{\psi_{\lambda}^{\brac{n}}} = \lim_{w \rightarrow 0} \radord{\func{\phi^{\brac{m}}}{z} \func{\psi_{\lambda}^{\brac{n}}}{w}} \ket{0}
\end{equation}
in powers of $z$, each term must have monodromy $-\theta_{\lambda} \pmod{1}$, so we are forced into an expansion of the form
\begin{equation}
\func{\phi^{\brac{m}}}{z} \ket{\psi_{\lambda}^{\brac{n}}} = \sum_{r \in \ZZ + \theta_{\lambda} - k/4} \phi^{\brac{m}}_r \ z^{-r-k/4} \ket{\psi_{\lambda}^{\brac{n}}}.
\end{equation}
In this way, the mode expansion of the simple current fields depend upon the state on which the modes act, as claimed.

The importance of the $\func{\alg{u}}{1}$-valued charges $\theta_{\lambda}^{\pair{m}{n}}$ is that they provide the additional information required to determine the largest value of $r$ in the above mode expansion such that $\phi^{\brac{m}}_r$ does not annihilate $\ket{\psi_{\lambda}^{\brac{n}}}$.  In other words, these charges determine the first non-trivial descendant states of $\ket{\psi_{\lambda}^{\brac{n}}}$ with respect to the algebra $\alg{A}_k$.  It is easily verified that these descendants are of the form
\begin{equation} \label{eqnFirstDesc}
\phi^{\brac{m}}_{\theta_{\lambda}^{\pair{m}{n}} - k/4} \ket{\psi_{\lambda}^{\brac{n}}} \qquad \text{(first descendants)},
\end{equation}
a result that will prove invaluable to us in what follows.  In particular, the vacuum has $\theta_0^{\pair{m}{0}} = 0$ for all $m$ ($n$ must be $0$), so its first $\alg{A}_k$-descendants are the zero-grade states $\ket{\phi^{\brac{m}}} = \phi^{\brac{m}}_{-k/4} \ket{0}$ (as we should expect).

It remains to determine the monodromy charge of descendant states.  Suppose then that $\ket{\psi}$ is a state of monodromy charge $\theta \in \ZZ / 2$ (this is true for zero-grade descendants, so we may suppose that it is true as part of an inductive procedure).  Then, $\phi^{\brac{m}}_r \ket{\psi}$ is defined if and only if $r \in \ZZ + \theta - k/4$, so by \eqnref{eqnAffPrimComm},
\begin{equation}
\phi^{\brac{m}}_r H_s \ket{\psi} = H_s \phi^{\brac{m}}_r \ket{\psi} - \brac{k-2m} \phi^{\brac{m}}_{r+s} \ket{\psi}
\end{equation}
must be defined for $r \in \ZZ + \theta - k/4$.  It follows that $H_s \ket{\psi}$ must also have monodromy charge $\theta$.  The same is true for $E_s \ket{\psi}$ and $F_s \ket{\psi}$, so we conclude that acting with affine modes does not change the monodromy charge (modulo $1$).

To determine the effect of acting with a $\phi$-mode on the monodromy charge, we apply the \gcr{} (\ref{eqnGeneralGCRs}) to the state $\ket{\psi}$.  For this to make sense, $s + \ell$ and $r - k/2 + \gamma + \ell$ must belong to $\ZZ + \theta - k/4$.  In other words, $r \in \ZZ + \theta + k/4$ and $s \in \ZZ + \theta - k/4$ (hence $r+s \in \ZZ$ as required on the right hand side of the \gcr{}).  It now follows from the term $\phi^{\brac{m}}_r \phi^{\brac{n}}_s \ket{\psi}$ that the monodromy charge of $\phi^{\brac{n}}_s \ket{\psi}$ must be $\theta + k/2$, hence we conclude that acting with $\phi$-modes changes the monodromy charge (modulo $1$) by $k/2$ (and by induction that the monodromy charge always takes values in $\ZZ / 2$).

We also mention that these results for descendant states can be derived directly from the definition in terms of the corresponding \ope{}.  We will usually only be interested in the charge (modulo $1$) of descendant states, so we will omit this demonstration.

\subsection{Highest Weight Modules} \label{secRepHWMs}

We define an $\alg{A}_k$-\hws{} to be an affine \hws{} $\ket{\psi_{\lambda}} \equiv \ket{\psi^{\brac{0}}_{\lambda}}$ such that a $\phi$-mode acting upon it cannot increase its $\func{\alg{sl}}{2}$-weight or lower its conformal dimension, just as the affine modes cannot in the affine case.  This requires
\begin{equation} \label{eqnHWSDef}
\phi^{\brac{m}}_r \ket{\psi^{\brac{0}}_{\lambda}} = 0 \qquad \text{for $r > 0$, or $r=0$ and $m < k/2$.}
\end{equation}
We now define an $\alg{A}_k$-Verma module in the usual way:  It is the module generated from an $\alg{A}_k$-\hws{} by the action of all elements of $\alg{A}_k$, modulo the algebra relations.

It remains to see whether the condition (\ref{eqnHWSDef}) imposes constraints on $\lambda$.  Being an affine \hws{}, the $\alg{A}_k$-\hws{} will obviously have a well-defined conformal dimension and $\func{\alg{sl}}{2}$-weight with $0 \leqslant \lambda \leqslant k$.  In addition, we may also associate to it the $\func{\alg{u}}{1}$-charges computed in \eqnref{eqnu1Charge}.  Since each $\alg{A}_k$-\hws{} must be of the form $\ket{\psi^{\brac{0}}_{\lambda}}$, its first non-vanishing descendants must be, by (\ref{eqnFirstDesc}), of the form ($m = 0, 1, \ldots, k$)
\begin{equation} \label{eqnFirstDescendant}
\phi^{\brac{m}}_{\theta_{\lambda}^{\pair{m}{0}} - k/4} \ket{\psi^{\brac{0}}_{\lambda}}, \quad \text{where} \quad \theta_{\lambda}^{\pair{m}{0}} = 
\begin{cases}
m - \lambda / 2 & \text{if $m \leqslant \lambda$,} \\
\lambda / 2 & \text{if $m \geqslant \lambda$.}
\end{cases}
\end{equation}
We observe that for $\lambda = k$, $\theta_k^{\pair{m}{0}} - k/4 = m - 3k/4$, which is positive for $m = k$.  In other words, we see that the affine \hws{} $\ket{\psi^{\brac{0}}_k}$ has a non-vanishing ``descendant'' of lower conformal dimension, hence we conclude that $\ket{\psi^{\brac{0}}_k}$ cannot be an $\alg{A}_k$-\hws{}.

This observation is easily generalised.  Since $\max \bigl\{ \theta_{\lambda}^{\pair{m}{0}} \colon m = 0, 1, \ldots, k \bigr\} =  \lambda / 2$, a sufficient condition for $\ket{\psi^{\brac{0}}_{\lambda}}$ to have a descendant of lower conformal dimension (hence not be an $\alg{A}_k$-\hws{}) is that $\lambda > k/2$.  If $\lambda = k/2$, we find that the first non-vanishing descendants (indexed by $m$ as above) have higher conformal dimensions than $\ket{\psi^{\brac{0}}_{k/2}}$ for $m < k/2$, and the conformal dimension is unchanged for $m \geqslant k/2$.  It is easy to check that in this latter case, the $\func{\alg{sl}}{2}$-weight of the descendants is less than or equal to that of $\ket{\psi^{\brac{0}}_{k/2}}$, so there is no contradiction to $\ket{\psi^{\brac{0}}_{k/2}}$ being an $\alg{A}_k$-\hws{}.

To summarise, we have shown that an affine \hws{} $\ket{\psi^{\brac{0}}_{\lambda}}$ is an $\alg{A}_k$-\hws{} if and only if $\lambda \leqslant k/2$.  The remaining affine \hwss{} must therefore appear as descendants of the genuine $\alg{A}_k$-\hwss{}.  The $\func{\alg{sl}}{2}$-weights and conformal dimensions of the affine \hwss{} suggest that the exact relationship is given by
\begin{equation} \label{eqnIdentifyAffHWSs}
\ket{\psi^{\brac{0}}_{k - \lambda}} = \phi^{\brac{\lambda}}_{\lambda / 2 - k/4} \ket{\psi^{\brac{0}}_{\lambda}},
\end{equation}
which is verified by directly showing that the right hand side is an affine \hws{}:
\begin{equation}
E_r \phi^{\brac{\lambda}}_{\lambda / 2 - k/4} \ket{\psi^{\brac{0}}_{\lambda}} = \acomm{E_r}{\phi^{\brac{\lambda}}_{\lambda / 2 - k/4}} \ket{\psi^{\brac{0}}_{\lambda}} = \bigl[ \lambda \brac{k+1 - \lambda} \bigr]^{1/2} \phi^{\brac{\lambda - 1}}_{\lambda / 2 + r - k/4} \ket{\psi^{\brac{0}}_{\lambda}} = 0,
\end{equation}
since $\theta_{\lambda}^{\pair{\lambda - 1}{0}} = \lambda / 2 - 1 < \lambda /2 + r$, for $r \geqslant 0$ (and similar calculations with $H_r$ and $F_r$, but $r > 0$).

Thus we see that a generic $\alg{A}_k$-Verma module contains at least\footnote{In fact, there are no more, as we shall prove in \propref{propTwoAffHWSs}.} two affine \hwss{}:  The $\alg{A}_k$-\hws{}, $\ket{\psi^{\brac{0}}_{\lambda}}$, and the $\alg{A}_k$-descendant $\ket{\psi^{\brac{0}}_{k - \lambda}}$.  It is necessary to consider separately the exceptional case where $k$ is even and $\lambda = k/2$.  Then these two affine \hwss{}, $\ket{\psi^{\brac{0}}_{k/2}}$ and $\ket{\widetilde{\psi}^{\brac{0}}_{k/2}}$ say, with
\begin{equation} \ket{\widetilde{\psi}^{\brac{0}}_{k/2}} = \phi^{\brac{k/2}}_0  \ket{\psi^{\brac{0}}_{k/2}},
\end{equation}
coincide (hence both would qualify as $\alg{A}_k$-\hwss{} according to the discussion above), and are related by the self-adjoint mode $\phi^{\brac{k/2}}_0$.  It is really a matter of convenience as to whether we should identify these two states or not.  We note that the simultaneous normalisation of all zero-grade states gives
\begin{equation}
1 = \braket{\widetilde{\psi}^{\brac{0}}_{k/2}}{\widetilde{\psi}^{\brac{0}}_{k/2}} = \bracket{\psi^{\brac{0}}_{k/2}}{\bigl( \phi^{\brac{k/2}}_0 \bigr)^2}{\psi^{\brac{0}}_{k/2}} \qquad \Rightarrow \qquad \bigl( \phi^{\brac{k/2}}_0 \bigr)^2 \ket{\psi^{\brac{0}}_{\lambda}} = \ket{\psi^{\brac{0}}_{\lambda}},
\end{equation}
so repeated application of $\phi^{\brac{k/2}}_0$ gives at most $2$ independent states.

If we choose to identify these two \hwss{}, we have to conclude that this exceptional $\alg{A}_k$-Verma module has but one affine \hws{}, and that $\phi^{\brac{k/2}}_0$ leaves this state invariant.  However, we will find it convenient (and not just for regularity of exposition) to not identify these states, and declare that $\ket{\psi^{\brac{0}}_{k/2}}$ is an $\alg{A}_k$-\hws{}, whilst $\ket{\widetilde{\psi}^{\brac{0}}_{k/2}}$ is not.  With this understanding, this exceptional module may be treated just like any other.

We point out that in the preceding discussion, we have only verified that \eqnref{eqnIdentifyAffHWSs} holds up to a multiplicative constant.  This reflects the fact that \hwss{} are only unique up to constant multipliers.  Recall from \secref{secBasics} that we have chosen to normalise all the affine \hwss{} (and their zero-grade descendants).  Identifying the affine \hwss{} as in \eqnref{eqnIdentifyAffHWSs} is consistent with this, because we may absorb any factors arising from this identification into the eigenvalues of the $\mathcal{S}_{m,n}$.  We now turn to an elaboration of this claim.

\subsection{$\mathcal{S}_{m,n}$-Eigenvalues} \label{secSEigs}

Let us consider the computation of the eigenvalues of the $\mathcal{S}_{m,n}$ on each $\alg{A}_k$-Verma module.  These eigenvalues can be calculated directly from the general commutation relations, but it can be rather tiresome to compute each of them in this way (there are $\bigl\lfloor \tfrac{1}{4} \brac{k+2}^2 \bigr\rfloor$ independent eigenvalues on each of the $\bigl\lceil \tfrac{1}{2} \brac{k+1} \bigr\rceil$ modules).  To illustrate this, and the method of computation, the simplest case is exhibited in \secref{secExamplek=2}.

However, recall that the purpose of the $\mathcal{S}_{m,n}$ is to absorb some unpleasant factors in the \opes{}, which derive from our insistence on simultaneously normalising the zero-grade states of a given module.  This is equivalent to independently choosing the inner-product on each affine module, and we have chosen it so that every affine \hws{} has unit norm.  In the extended theory, affine \hwss{} are paired in the $\alg{A}_k$-Verma modules, so the algebra could potentially interfere with our ability to independently choose the inner-products.  As mentioned above, however, the $\mathcal{S}_{m,n}$ are able to absorb the factor that the algebra dictates should relate the two inner-products, assuming \eqnref{eqnIdentifyAffHWSs}, thereby restoring our ability to choose to normalise every zero-grade state.  This factor will clearly depend upon the module, but \emph{not} upon $m$ and $n$.

The upshot of this is that in the formalism we have developed, the eigenvalue of $\mathcal{S}_{m,n}$ on some $\alg{A}_k$-Verma module is the product of its eigenvalue on $\ket{0}$ (given by \eqnref{eqnSValues}) and a factor independent of $m$ and $n$.  For each module, this latter factor may therefore be computed from the eigenvalue of a \emph{single} $\mathcal{S}_{m,n}$.  A convenient eigenvalue on $\ket{\psi_{\lambda}} \equiv \ket{\psi^{\brac{0}}_{\lambda}}$ ($\lambda \leqslant k/2$) for this computation would be that of $\mathcal{S}_{k-\lambda,\lambda}$, because
\begin{equation}
1 = \braket{\psi_{k-\lambda}}{\psi_{k-\lambda}} = \bracket{\psi_{\lambda}}{\phi^{\brac{k-\lambda}}_{k/4 - \lambda/2} \phi^{\brac{\lambda}}_{\lambda/2 - k/4}}{\psi_{\lambda}}
\gcreq{\lambda + 1} \sum_{j=0}^{\lambda} \binom{\lambda / 2}{\lambda - j} \bracket{\psi_{\lambda}}{\mathcal{S}_{k-\lambda,\lambda} A^{\brac{k - \lambda , \lambda ; j}}_0}{\psi_{\lambda}}.
\end{equation}
The $A^{\brac{k - \lambda , \lambda ; j}}_0$ are $\func{\alg{sl}}{2}$-weightless, dimensionless, normally-ordered combinations of affine modes, hence their action on an affine \hws{} is by scalar multiplication.  Evaluating this action then allows us to determine the eigenvalue of $\mathcal{S}_{k-\lambda,\lambda}$, hence the eigenvalues of all of the $\mathcal{S}_{m,n}$.

We illustrate this with the first few cases.  If $\lambda = 0$ ($\ket{\psi_{\lambda}} = \ket{0}$), then $\mathcal{S}_{k,0}$ is easily computed to have eigenvalue $1$.  The eigenvalues of a general $\mathcal{S}_{m,n}$ on the vacuum are therefore given by \eqnref{eqnSValues} (as they must be).  The first non-trivial case is $\lambda = 1$ which gives
\begin{equation}
1 \gcreq{2} \bracket{\psi_1}{\mathcal{S}_{k-1,1} \sqbrac{\frac{1}{2} - \frac{k-2}{2k} H_0}}{\psi_1} = \frac{1}{k} \bracket{\psi_1}{\mathcal{S}_{k-1,1}}{\psi_1},
\end{equation}
hence eigenvalue $k$.  The eigenvalues of the $\mathcal{S}_{m,n}$ on $\ket{\psi_1}$ are therefore $k$ times the values given by \eqnref{eqnSValues}.  For $\lambda = 2$ we get
\begin{align}
1 &\gcreq{3} \bracket{\psi_2}{\mathcal{S}_{k-2,2} \sqbrac{\frac{4-k}{2k} H_0 + \frac{k+2}{6} L_0 - \frac{4-k}{4k} H_0 + \brac{\frac{1}{12} - \frac{k-2}{k \brac{k-1}}} \Delta^0_0}}{\psi_2} \notag \\
&= \frac{2}{k \brac{k-1}} \bracket{\psi_2}{\mathcal{S}_{k-2,2}}{\psi_2},
\end{align}
where we note that
\begin{equation}
\Delta^0_0 \ket{\psi_2} = \sum_r \normord{H_r H_{-r} - E_r F_{-r} - F_r E_{-r}} \ket{\psi_2} = \brac{H_0^2 - H_0} \ket{\psi_2} = 2 \ket{\psi_2}
\end{equation}
and $\brac{\partial H}_0 = -H_0$.  The $\mathcal{S}_{m,n}$-eigenvalues are therefore $\binom{k}{2}$ times the eigenvalues given in \eqnref{eqnSValues}.

It appears likely from these computations that this ratio will be $\binom{k}{\lambda}$ in general, but it is difficult to prove this using the above method.  The problem is essentially that we have to use the \gcr{} of order $\gamma = \lambda + 1$, thereby needing to compute the $A^{\brac{k - \lambda , \lambda ; j}}_0$ for $j$ up to $\lambda$, a rather non-trivial task when $\lambda$ is large.  This non-triviality derives from the large number of terms contributing to the right-hand-side of the \gcr{} with $m + n = k$ when $\gamma = \lambda + 1$ is large.  Such general computations would be far easier if we were only required to use a \gcr{} with a single (known) term on the right-hand-side.  Remarkably, this can indeed be arranged.

\begin{proposition} \label{propSEigs}
The eigenvalue of $\mathcal{S}_{m,n}$ on $\ket{\psi_{\lambda}}$ is $\binom{k}{\lambda}$ times its eigenvalue on $\ket{0}$.
\end{proposition}
\begin{proof}
By iterating \eqnref{eqnSU2Normalisation}, we may write
\begin{equation}
1 = \braket{\psi_{\lambda}}{\psi_{\lambda}} = \frac{1}{\lambda !} \bracket{\psi^{\brac{0}}_{\lambda}}{E_0^{\lambda}}{\psi^{\brac{\lambda}}_{\lambda}}.
\end{equation}
Since $\psi^{\brac{\lambda}}_{\lambda}$ is a zero-grade state, we have
\begin{equation}
E_0^{\lambda} \ket{\psi^{\brac{\lambda}}_{\lambda}} = \sum_{\substack{r_1 , \ldots , r_{\lambda} \in \ZZ \\ r_1 + \ldots + r_{\lambda} = 0}} \normord{E_{r_1} \cdots E_{r_{\lambda}}} \ket{\psi^{\brac{\lambda}}_{\lambda}},
\end{equation}
where the normal-ordering is nested to the right as usual.  But, this is (up to an $\mathcal{S}_{m,n}$ factor) precisely the right-hand-side of the \gcr{} with $\gamma = \lambda + 1$, $m + n = k - \lambda$ and $r + s = 0$.  We choose $n = 0$ and $s = \lambda / 2 - k / 4$ (in accord with the monodromy charge of $\ket{\psi^{\brac{\lambda}}_{\lambda}}$):
\begin{multline}
\sum_{\ell = 0}^{\infty} \binom{\ell - k/2 + \lambda}{\ell} \sqbrac{\phi^{\brac{k - \lambda}}_{k/4 - \lambda / 2 - \ell} \phi^{\brac{0}}_{\lambda / 2 - k/4 + \ell} + \phi^{\brac{0}}_{k/4 - \lambda / 2 - 1 - \ell} \phi^{\brac{k - \lambda}}_{\lambda / 2 - k/4 + 1 + \ell}} \\*
\gcreq{\lambda + 1} \mathcal{S}_{k - \lambda , 0} \sum_{\substack{r_1 , \ldots , r_{\lambda} \in \ZZ \\ r_1 + \ldots + r_{\lambda} = 0}} \normord{E_{r_1} \cdots E_{r_{\lambda}}}.
\end{multline}
Applying this to $\ket{\psi^{\brac{\lambda}}_{\lambda}}$, almost all of the terms vanish because $\theta_{\lambda}^{\pair{m}{\lambda}} = \lambda / 2$ when $m \leqslant k - \lambda$ (by \eqnref{eqnu1Charge}).  We therefore conclude that
\begin{align}
1 &= \frac{1}{\lambda !} \bracket{\psi^{\brac{0}}_{\lambda}}{\sum_{\substack{r_1 , \ldots , r_{\lambda} \in \ZZ \\ r_1 + \ldots + r_{\lambda} = 0}} \normord{E_{r_1} \cdots E_{r_{\lambda}}}}{\psi^{\brac{\lambda}}_{\lambda}} \gcreq{\lambda + 1} \frac{1}{\lambda !} \bracket{\psi^{\brac{0}}_{\lambda}}{\mathcal{S}_{k - \lambda , 0}^{-1} \phi^{\brac{k - \lambda}}_{k/4 - \lambda / 2} \phi^{\brac{0}}_{\lambda / 2 - k/4}}{\psi^{\brac{\lambda}}_{\lambda}} \notag \\
&= \frac{1}{\lambda !} \bracket{\psi^{\brac{0}}_{k - \lambda}}{\mathcal{S}_{k - \lambda , 0}^{-1} \phi^{\brac{0}}_{\lambda / 2 - k/4}}{\psi^{\brac{\lambda}}_{\lambda}},
\end{align}
by (the adjoint of) \eqnref{eqnIdentifyAffHWSs}.

Consider now the combination $\phi^{\brac{0}}_{\lambda / 2 - k/4} \ket{\psi^{\brac{\lambda}}_{\lambda}}$.  \eqnref{eqnSU2Normalisation} allows us to ``exchange'' the $\func{\alg{sl}}{2}$-weight between the mode and the state in the following manner:
\begin{align}
\phi^{\brac{0}}_{\lambda / 2 - k/4} \ket{\psi^{\brac{\lambda}}_{\lambda}} &= \frac{1}{\lambda^{1/2}} \phi^{\brac{0}}_{\lambda / 2 - k/4} F_0 \ket{\psi^{\brac{\lambda - 1}}_{\lambda}} \notag \\
&= \sqbrac{\frac{k}{\lambda}}^{1/2} \phi^{\brac{1}}_{\lambda / 2 - k/4} \ket{\psi^{\brac{\lambda - 1}}_{\lambda}} - \frac{1}{\lambda^{1/2}} F_0 \phi^{\brac{0}}_{\lambda / 2 - k/4} \ket{\psi^{\brac{\lambda - 1}}_{\lambda}}.
\end{align}
The $F_0$ at the left of the last term will annihilate $\bra{\psi^{\brac{0}}_{k - \lambda}}$, so iterating this exchange gives
\begin{equation}
1 = \frac{1}{\lambda !} \binom{k}{\lambda}^{1/2} \bracket{\psi^{\brac{0}}_{k - \lambda}}{\mathcal{S}_{k - \lambda , 0}^{-1} \phi^{\brac{\lambda}}_{\lambda / 2 - k/4}}{\psi^{\brac{0}}_{\lambda}} = \frac{1}{\lambda !} \binom{k}{\lambda}^{1/2} \bracket{\psi^{\brac{0}}_{k - \lambda}}{\mathcal{S}_{k - \lambda , 0}^{-1}}{\psi^{\brac{0}}_{k - \lambda}}.
\end{equation}
The eigenvalues of $\mathcal{S}_{k - \lambda , 0}$ on $\ket{\psi_{k - \lambda}}$ and on the vacuum (see \eqnref{eqnSValues}) are therefore given by
\begin{equation}
\frac{1}{\lambda !} \binom{k}{\lambda}^{1/2} \qquad \text{and} \qquad \frac{\brac{k - \lambda} !}{k !} \binom{k}{\lambda}^{1/2}
\end{equation}
respectively, whence the result.
\end{proof}

\section{Examples} \label{secExamples}

\subsection{The $\func{\affine{\alg{sl}}}{2}_2$ Extended Algebra} \label{secExamplek=2}

We consider the case $k=2$, so there are three affine primaries:  The identity field, a field $\func{\psi_1}{z}$ of conformal dimension $\frac{3}{16}$, and a simple current $\func{\phi}{z}$ of conformal dimension $\frac{1}{2}$.  $\func{\psi_1}{z}$ gives rise to two zero-grade descendant fields, $\func{\psi_1}{z} \equiv \func{\psi^{\brac{0}}_1}{z}$ and $\func{\psi^{\brac{1}}_1}{z}$ (of respective weights $1$ and $-1$), whereas the simple current gives rise to three:  $\func{\phi}{z} \equiv \func{\phi^{\brac{0}}}{z}$, $\func{\phi^{\brac{1}}}{z}$ and $\func{\phi^{\brac{2}}}{z}$ (of respective weights $2$, $0$, and $-2$).  The operator product expansions of the latter take the form:
\begin{align}
\radord{\func{\phi^{\brac{0}}}{z} \func{\phi^{\brac{0}}}{w}} &= \mathcal{S}_{0,0} \normord{\func{E}{w} \func{E}{w}} \brac{z-w} + \ldots \notag \\
\radord{\func{\phi^{\brac{0}}}{z} \func{\phi^{\brac{1}}}{w}} &= \mathcal{S}_{0,1} \sqbrac{\func{E}{w} + \brac{\tfrac{1}{2} \func{\partial E}{w} + \tfrac{1}{4} \func{\Delta^+}{w}} \brac{z-w} + \ldots} \notag \\
\radord{\func{\phi^{\brac{1}}}{z} \func{\phi^{\brac{0}}}{w}} &= \mathcal{S}_{1,0} \sqbrac{\func{E}{w} + \brac{\tfrac{1}{2} \func{\partial E}{w} - \tfrac{1}{4} \func{\Delta^+}{w}} \brac{z-w} + \ldots} \notag \\
\radord{\func{\phi^{\brac{0}}}{z} \func{\phi^{\brac{2}}}{w}} &= \mathcal{S}_{0,2} \sqbrac{\frac{1}{z-w} + \tfrac{1}{2} \func{H}{w} + \brac{\tfrac{2}{3} \func{T}{w} + \tfrac{1}{4} \func{\partial H}{w} + \tfrac{1}{12} \func{\Delta^0}{w}} \brac{z-w} + \ldots} \notag \\
\radord{\func{\phi^{\brac{1}}}{z} \func{\phi^{\brac{1}}}{w}} &= \mathcal{S}_{1,1} \sqbrac{\frac{1}{z-w} + \brac{\tfrac{2}{3} \func{T}{w} - \tfrac{1}{6} \func{\Delta^0}{w}} \brac{z-w} + \ldots} \\
\radord{\func{\phi^{\brac{2}}}{z} \func{\phi^{\brac{0}}}{w}} &= \mathcal{S}_{2,0} \sqbrac{\frac{1}{z-w} - \tfrac{1}{2} \func{H}{w} + \brac{\tfrac{2}{3} \func{T}{w} - \tfrac{1}{4} \func{\partial H}{w} + \tfrac{1}{12} \func{\Delta^0}{w}} \brac{z-w} + \ldots} \notag \\
\radord{\func{\phi^{\brac{1}}}{z} \func{\phi^{\brac{2}}}{w}} &= \mathcal{S}_{1,2} \sqbrac{\func{F}{w} + \brac{\tfrac{1}{2} \func{\partial F}{w} + \tfrac{1}{4} \func{\Delta^-}{w}} \brac{z-w} + \ldots} \notag \\
\radord{\func{\phi^{\brac{2}}}{z} \func{\phi^{\brac{1}}}{w}} &= \mathcal{S}_{2,1} \sqbrac{\func{F}{w} + \brac{\tfrac{1}{2} \func{\partial F}{w} - \tfrac{1}{4} \func{\Delta^-}{w}} \brac{z-w} + \ldots} \notag \\
\radord{\func{\phi^{\brac{2}}}{z} \func{\phi^{\brac{2}}}{w}} &= \mathcal{S}_{2,2} \normord{\func{F}{w} \func{F}{w}} \brac{z-w} + \ldots \notag
\end{align}
We have explicitly listed both $\radord{\func{\phi^{\brac{m}}}{z} \func{\phi^{\brac{n}}}{w}}$ and $\radord{\func{\phi^{\brac{n}}}{z} \func{\phi^{\brac{m}}}{w}}$ in order to illustrate the commutativity given by \eqnref{eqnROEComm} (with $\mu_{m,n} = 1$):
\begin{equation} \label{eqnOPECommk=2}
\radord{\func{\phi^{\brac{m}}}{z} \func{\phi^{\brac{n}}}{w}} = \brac{-1}^{m+n+1} \radord{\func{\phi^{\brac{n}}}{w} \func{\phi^{\brac{m}}}{z}}.
\end{equation}
We also note that the \gcrs{} with $\gamma = 1$ are the super-commutation relations
\begin{equation} \label{eqnSuperCommk=2}
\phi^{\brac{m}}_r \phi^{\brac{n}}_s + \brac{-1}^{m+n} \phi^{\brac{n}}_s \phi^{\brac{m}}_r \gcreq{1} \mathcal{S}_{m,n} \delta_{m+n , 2} \delta_{r+s , 0}.
\end{equation}

From \eqnref{eqnSValues}, the eigenvalues of the $\mathcal{S}_{m,n}$ on the vacuum $\alg{A}_2$-module are given (in matrix form with rows and columns indexed by $0$, $1$ and $2$) by
\begin{equation} \label{eqnSVacEigsk=2}
\mathcal{S} \ket{0} = 
\begin{pmatrix}
1/2 & 1/\sqrt{2} & 1 \\
1/\sqrt{2} & 1 & 1/\sqrt{2} \\
1 & 1/\sqrt{2} & 1/2
\end{pmatrix}
\ket{0}.
\end{equation}
We determine the eigenvalues on $\ket{\psi_1}$, to compare with (and illustrate) the discussion of \secref{secSEigs}.  The easiest determination is
\begin{equation}
1 = \braket{\widetilde{\psi}_1}{\widetilde{\psi}_1} = \bracket{\psi_1}{\phi^{\brac{1}}_0 \phi^{\brac{1}}_0}{\psi_1} \gcreq{1} \frac{1}{2} \bracket{\psi_1}{\mathcal{S}_{1,1}}{\psi_1},
\end{equation}
hence $\mathcal{S}_{1,1} \ket{\psi_1} = 2 \ket{\psi_1}$.

The computation of the other eigenvalues provides a convenient illustration of the use of \gcrs{}.  First, note that $\gamma = 1$ gives
\begin{equation}
\bracket{\psi_1}{\phi^{\brac{0}}_0 \phi^{\brac{1}}_0 F_0}{\psi_1} \gcreq{1} \bracket{\psi_1}{\phi^{\brac{1}}_0 \phi^{\brac{0}}_0 F_0}{\psi_1} = \sqrt{2} \bracket{\psi_1}{\phi^{\brac{1}}_0 \phi^{\brac{1}}_0}{\psi_1} = \sqrt{2},
\end{equation}
since $\phi^{\brac{0}}_0$ annihilates $\ket{\psi_1}$, by \eqnref{eqnFirstDescendant}.  Alternatively,
\begin{equation}
\bracket{\psi_1}{\phi^{\brac{0}}_0 \phi^{\brac{1}}_0 F_0}{\psi_1} \gcreq{2} \bracket{\psi_1}{\mathcal{S}_{0,1} E_0 F_0}{\psi_1} = \bracket{\psi_1}{\mathcal{S}_{0,1}}{\psi_1},
\end{equation}
and
\begin{align}
\bracket{\psi_1}{\phi^{\brac{0}}_0 \phi^{\brac{1}}_0 F_0}{\psi_1} &= \sqrt{2} \bracket{\psi_1}{\phi^{\brac{0}}_0 \phi^{\brac{2}}_0}{\psi_1} - \bracket{\psi_1}{\phi^{\brac{0}}_0 F_0 \phi^{\brac{1}}_0}{\psi_1} \notag \\
&\gcreq{1} \sqrt{2} \bracket{\psi_1}{\mathcal{S}_{0,2}}{\psi_1} - \sqrt{2} \bracket{\psi_1}{\phi^{\brac{1}}_0 \phi^{\brac{1}}_0}{\psi_1} = \sqrt{2} \Bigl[ \bracket{\psi_1}{\mathcal{S}_{0,2}}{\psi_1} - 1 \Bigr],
\end{align}
from which we get $\mathcal{S}_{0,1} \ket{\psi_1} = \sqrt{2} \ket{\psi_1}$ and $\mathcal{S}_{0,2} \ket{\psi_1} = 2 \ket{\psi_1}$.  Finally, the eigenvalue of $\mathcal{S}_{0,0}$ may be obtained by comparing
\begin{equation}
\bracket{\psi_1}{F_1 \phi^{\brac{0}}_{-1} \phi^{\brac{0}}_1 F_{-1}}{\psi_1} = 2 \bracket{\psi_1}{\phi^{\brac{1}}_0 \phi^{\brac{1}}_0}{\psi_1} = 2,
\end{equation}
and
\begin{equation}
\bracket{\psi_1}{F_1 \phi^{\brac{0}}_{-1} \phi^{\brac{0}}_1 F_{-1}}{\psi_1} \gcreq{3} \frac{1}{2} \bracket{\psi_1}{\mathcal{S}_{0,0} F_1 \brac{E_0^2 + 2 E_{-1} E_1} F_{-1}}{\psi_1} = 2 \bracket{\psi_1}{\mathcal{S}_{0,0}}{\psi_1},
\end{equation}
whence $\mathcal{S}_{0,0} \ket{\psi_1} = \ket{\psi_1}$, and
\begin{equation} \label{eqnSOtherEigsk=2}
\mathcal{S} \ket{\psi_1} = 
\begin{pmatrix}
1 & \sqrt{2} & 2 \\
\sqrt{2} & 2 & \sqrt{2} \\
2 & \sqrt{2} & 1
\end{pmatrix}
\ket{\psi_1}.
\end{equation}
The eigenvalues of the $\mathcal{S}_{m,n}$ on $\ket{\psi_1}$ are therefore precisely double the corresponding eigenvalues on $\ket{0}$, explicitly confirming the discussion in \secref{secSEigs}.

We conclude this example by showing how $\alg{A}_2$ decomposes into three copies of the extended symmetry algebra of the Ising model $\MinMod{3}{4}$.  By this, we mean the Ising model extended by its own simple current $\phi_{2,1}$, which effectively describes a free fermion.  We define two fermions by
\begin{equation}
\func{\chi^+}{z} = \frac{1}{\sqrt{2}} \sqbrac{\func{\phi^{\brac{0}}}{z} + \func{\phi^{\brac{2}}}{z}} \quad \text{and} \quad \func{\chi^-}{z} = \frac{- \ii}{\sqrt{2}} \sqbrac{\func{\phi^{\brac{0}}}{z} - \func{\phi^{\brac{2}}}{z}}.
\end{equation}
Then, the \opes{} give:
\begin{align}
\radord{\func{\chi^+}{z} \func{\chi^+}{w}} &= \mathcal{S}_{0,2} \sqbrac{\frac{1}{z-w} + \brac{\tfrac{2}{3} \func{T}{w} + \tfrac{1}{12} \func{\Delta^0}{w} + \tfrac{1}{4} \func{\normord{E E + FF}}{w}} \brac{z-w} + \ldots} \notag \\
\radord{\func{\chi^+}{z} \func{\chi^-}{w}} &= \mathcal{S}_{0,2} \frac{\ii}{2} \sqbrac{\func{H}{w} + \tfrac{1}{2} \brac{\func{\partial H}{w} - \func{\normord{E E - FF}}{w}} \brac{z-w} + \ldots} \\
\radord{\func{\chi^-}{z} \func{\chi^-}{w}} &= \mathcal{S}_{0,2} \sqbrac{\frac{1}{z-w} + \brac{\tfrac{2}{3} \func{T}{w} + \tfrac{1}{12} \func{\Delta^0}{w} - \tfrac{1}{4} \func{\normord{E E + FF}}{w}} \brac{z-w} + \ldots}, \notag
\end{align}
where we have used the fact that $\mathcal{S}_{0,0} = \tfrac{1}{2} \mathcal{S}_{0,2}$.

The third fermion field should be $\func{\phi^{\brac{1}}}{z}$, but we note that this field \emph{commutes} with the other fermions, by \eqnref{eqnOPECommk=2}, instead of anticommuting with them.  The remedy is to introduce an operator $\mathcal{F}$ into the fermion definitions so as to restore anticommutativity.  Recall from \secref{secAlgPrelim} that when $k$ is even, we may consistently choose a different adjoint in the extended theory which results in the affine modes $E$ and $F$ commuting with the $\phi$-modes, rather than anticommuting (this is referred to as choice (\ref{ChoiceBoson}) in \secref{secAlgPrelim}).  The equivalence of the theory defined by this adjoint and the adjoint we have used throughout was given by such an $\mathcal{F}$ in \eqnref{eqnFDef}, whose properties were summarised at the end of \secref{secAlgPrelim}.

Redefining the first two fermions as $\func{\chi^+}{z} \rightarrow \func{\chi^+}{z} \mathcal{F}$ and $\func{\chi^-}{z} \rightarrow \func{\chi^-}{z} \mathcal{F}$ has no effect on their \opes{}.  We now define the third fermion to be $\func{\chi^0}{z} = \ii \func{\phi^{\brac{1}}}{z} \mathcal{F}$, so that the three fermions are all mutually anticommutative.  The remaining \opes{} then become
\begin{align}
\radord{\func{\chi^+}{z} \func{\chi^0}{w}} &= \mathcal{S}_{0,2} \frac{- \ii}{2} \sqbrac{\func{\brac{E+F}}{w} + \brac{\tfrac{1}{2} \func{\partial \brac{E+F}}{w} + \tfrac{1}{4} \func{\brac{\Delta^+ - \Delta^-}}{w}} \brac{z-w} + \ldots} \notag \\
\radord{\func{\chi^0}{z} \func{\chi^0}{w}} &= \mathcal{S}_{0,2} \sqbrac{\frac{1}{z-w} + \brac{\tfrac{2}{3} \func{T}{w} - \tfrac{1}{6} \func{\Delta^0}{w}} \brac{z-w} + \ldots} \\
\radord{\func{\chi^-}{z} \func{\chi^0}{w}} &= \mathcal{S}_{0,2} \frac{-1}{2} \sqbrac{\func{\brac{E-F}}{w} + \brac{\tfrac{1}{2} \func{\partial \brac{E-F}}{w} + \tfrac{1}{4} \func{\brac{\Delta^+ + \Delta^-}}{w}} \brac{z-w} + \ldots}, \notag
\end{align}
using $\mathcal{S}_{1,1} = \mathcal{S}_{0,2}$ and $\mathcal{S}_{0,1} = \tfrac{1}{\sqrt{2}} \mathcal{S}_{0,2}$.  Note that the \opes{} of distinct fermions are regular.

These \opes{} should be compared with the \ope{} of the simple current\footnote{Quantities with bars such as these will always be reserved for those relating to the extended Ising model (fermion) theory.  They do not refer to antiholomorphic sectors.} $\func{\overline{\chi}}{z}$ of the Ising model \cite{DiFCon97,RidMin06}:
\begin{equation}
\radord{\func{\overline{\chi}}{z} \func{\overline{\chi}}{w}} = \overline{\mathcal{S}} \sqbrac{\frac{1}{z-w} + 2 \func{\overline{T}}{w} + \ldots}.
\end{equation}
Here, $\overline{\mathcal{S}}$ is a scalar operator taking (when the Virasoro \hwss{} are simultaneously normalised) eigenvalue $1$ on the Ising model vacuum and $2$ on the \hws{} of conformal dimension $\tfrac{1}{16}$ (so we identify it with $\mathcal{S}_{0,2} = \mathcal{S}_{1,1}$), and $\func{\overline{T}}{w}$ denotes the Ising model Virasoro field.  It follows that the $\alg{A}_2$-Virasoro field decomposes into three Ising model Virasoro fields, $\func{T}{z} = \func{\overline{T}^+}{z} + \func{\overline{T}^0}{z} + \func{\overline{T}^-}{z}$, where
\begin{align}
\func{\overline{T}^+}{z} &= \tfrac{1}{3} \func{T}{z} + \tfrac{1}{24} \func{\Delta^0}{z} + \tfrac{1}{8} \func{\normord{EE + FF}}{z} = \tfrac{1}{16} \func{\normord{HH}}{z} + \tfrac{1}{8} \func{\normord{EE + FF}}{z}, \notag \\
\func{\overline{T}^0}{z} &= \tfrac{1}{3} \func{T}{z} - \tfrac{1}{12} \func{\Delta^0}{z} = \tfrac{-1}{16} \func{\normord{HH}}{z} + \tfrac{1}{8} \func{\normord{EF + FE}}{z}, \\
\text{and} \quad \func{\overline{T}^-}{z} &= \tfrac{1}{3} \func{T}{z} + \tfrac{1}{24} \func{\Delta^0}{z} - \tfrac{1}{8} \func{\normord{EE + FF}}{z} = \tfrac{1}{16} \func{\normord{HH}}{z} - \tfrac{1}{8} \func{\normord{EE + FF}}{z}. \notag
\end{align}
One may explicitly check (we used \textsc{OPEdefs} \cite{ThiOPE91} in \textsc{Mathematica}) that these are indeed Virasoro fields of central charge $\tfrac{1}{2}$, and that the \opes{} between distinct Virasoro fields are regular\footnote{Actually, \textsc{OPEdefs} reports that these \opes{} all have the form
\begin{equation}
\radord{\func{\overline{T}^i}{z} \func{\overline{T}^j}{w}} = \pm \frac{1}{8} \frac{\func{\normord{E H E}}{w} - \func{\normord{F H F}}{w}}{z-w} + \ldots \qquad \text{($i, j \in \set{+,0,-} , i \neq j$).}
\end{equation}
However, it is easy to check that the two composite fields appearing above are actually null.}.

Denoting the extended algebra of the Ising model by $\overline{\alg{A}}$, we have therefore explicitly demonstrated the isomorphism
\begin{equation}
\alg{A}_2 \cong \overline{\alg{A}} \otimes \overline{\alg{A}} \otimes \overline{\alg{A}}.
\end{equation}
We note that this isomorphism preserves the adjoints of the theories (the fermions constructed above are self-adjoint), hence that we have constructed an isomorphism of $L_0$-graded $^*$-algebras (the grading by $\func{\alg{sl}}{2}$-weight is lost on the fermions).  Moreover, it is not hard to verify that the $\alg{A}_2$-\hwss{} $\ket{0}$ and $\ket{\psi_1}$ decompose under this isomorphism as
\begin{equation}
\ket{0} \longleftrightarrow \ket{\overline{0}} \otimes \ket{\overline{0}} \otimes \ket{\overline{0}} \qquad \text{and} \qquad \ket{\psi_1} \longleftrightarrow \ket{\overline{\tfrac{1}{16}}} \otimes \ket{\overline{\tfrac{1}{16}}} \otimes \ket{\overline{\tfrac{1}{16}}},
\end{equation}
respectively, where $\ket{\overline{h}}$ denotes a $\overline{\alg{A}}$-\hws{} of conformal dimension $h$ (the affine \hws{} $\ket{\phi}$ does not even correspond to an eigenstate of the Virasoro modes $\overline{L}_0^{\pm}$ corresponding to the fields $\func{\overline{T}^{\pm}}{z}$).  This isomorphism of algebras suggests that there are corresponding isomorphisms of modules, hence character identities.  We will study this in \secDref{secSingVects}{secCharacters}.

\subsection{The $\func{\affine{\alg{sl}}}{2}_4$ Extended Algebra} \label{secExamplek=4}

When $k = 4$, there are three $\alg{A}_4$-\hwss{}, $\ket{0}$, $\ket{\psi_1}$ and $\ket{\psi_2}$, and the simple current has five component fields $\func{\phi^{\brac{m}}}{z}$, $m = 0, \ldots, 4$, of conformal dimension $1$.  The $\mathcal{S}_{m,n}$ are given by \eqnref{eqnSValues} as
\begin{equation}
\mathcal{S} \ket{0} = 
\begin{pmatrix}
1/24 & 1/12 & \sqrt{6}/12 & 1/2 & 1 \\
1/12 & 1/4 & \sqrt{6}/4 & 1 & 1/2 \\
\sqrt{6}/12 & \sqrt{6}/4 & 1 & \sqrt{6}/4 & \sqrt{6}/12 \\
1/2 & 1 & \sqrt{6}/4 & 1/4 & 1/12 \\
1 & 1/2 & \sqrt{6}/12 & 1/12 & 1/24
\end{pmatrix}
\ket{0},
\end{equation}
and \secref{secSEigs} gives the corresponding eigenvalues on $\ket{\psi_1}$ and $\ket{\psi_2}$ as $4$ and $6$ times these, respectively.  The simple current component fields are supercommutative in the sense that
\begin{equation} \label{eqnOPECommk=4}
\radord{\func{\phi^{\brac{m}}}{z} \func{\phi^{\brac{n}}}{w}} = \brac{-1}^{m+n} \radord{\func{\phi^{\brac{n}}}{w} \func{\phi^{\brac{m}}}{z}},
\end{equation}
and the non-zero \opes{} (to second order) are
\begin{align}
\radord{\func{\phi^{\brac{1}}}{z} \func{\phi^{\brac{1}}}{w}} &= \frac{1}{4} \mathcal{S}_{2,2} \sqbrac{\func{\normord{EE}}{w} + \ldots} \notag \\*
\radord{\func{\phi^{\brac{0}}}{z} \func{\phi^{\brac{2}}}{w}} &= \frac{\sqrt{6}}{12} \mathcal{S}_{2,2} \sqbrac{\func{\normord{EE}}{w} + \ldots} \notag \\
\radord{\func{\phi^{\brac{2}}}{z} \func{\phi^{\brac{4}}}{w}} &= \frac{\sqrt{6}}{12} \mathcal{S}_{2,2} \sqbrac{\func{\normord{FF}}{w} + \ldots} \notag \\*
\radord{\func{\phi^{\brac{3}}}{z} \func{\phi^{\brac{3}}}{w}} &= \frac{1}{4} \mathcal{S}_{2,2} \sqbrac{\func{\normord{FF}}{w} + \ldots} \notag \\
\radord{\func{\phi^{\brac{0}}}{z} \func{\phi^{\brac{3}}}{w}} &= \frac{1}{2} \mathcal{S}_{2,2} \sqbrac{\frac{\func{E}{w}}{z-w} + \tfrac{1}{2} \func{\partial E}{w} + \tfrac{1}{4} \func{\Delta^+}{w} + \ldots} \notag \\*
\radord{\func{\phi^{\brac{1}}}{z} \func{\phi^{\brac{2}}}{w}} &= \frac{\sqrt{6}}{4} \mathcal{S}_{2,2} \sqbrac{\frac{\func{E}{w}}{z-w} + \tfrac{1}{2} \func{\partial E}{w} + \tfrac{1}{12} \func{\Delta^+}{w} + \ldots} \\
\radord{\func{\phi^{\brac{2}}}{z} \func{\phi^{\brac{3}}}{w}} &= \frac{\sqrt{6}}{4} \mathcal{S}_{2,2} \sqbrac{\frac{\func{F}{w}}{z-w} + \tfrac{1}{2} \func{\partial F}{w} + \tfrac{1}{12} \func{\Delta^-}{w} + \ldots} \notag \\*
\radord{\func{\phi^{\brac{1}}}{z} \func{\phi^{\brac{4}}}{w}} &= \frac{1}{2} \mathcal{S}_{2,2} \sqbrac{\frac{\func{F}{w}}{z-w} + \tfrac{1}{2} \func{\partial F}{w} + \tfrac{1}{4} \func{\Delta^-}{w} + \ldots} \notag \\
\radord{\func{\phi^{\brac{0}}}{z} \func{\phi^{\brac{4}}}{w}} &= \mathcal{S}_{2,2} \sqbrac{\frac{1}{\brac{z-w}^2} + \frac{\func{H}{w}}{2 \brac{z-w}} + \func{T}{w} + \tfrac{1}{4} \func{\partial H}{w} + \tfrac{1}{12} \func{\Delta^0}{w} + \ldots} \notag \\*
\radord{\func{\phi^{\brac{1}}}{z} \func{\phi^{\brac{3}}}{w}} &= \mathcal{S}_{2,2} \sqbrac{\frac{1}{\brac{z-w}^2} + \frac{\func{H}{w}}{4 \brac{z-w}} + \func{T}{w} + \tfrac{1}{8} \func{\partial H}{w} - \tfrac{1}{24} \func{\Delta^0}{w} + \ldots} \notag \\*
\radord{\func{\phi^{\brac{2}}}{z} \func{\phi^{\brac{2}}}{w}} &= \mathcal{S}_{2,2} \sqbrac{\frac{1}{\brac{z-w}^2} \mspace{5.7mu} \phantom{ + \frac{\func{H}{w}}{4 \brac{z-w}}} + \func{T}{w} \mspace{1.8mu} \phantom{ + \tfrac{1}{8} \func{\partial H}{w}} - \tfrac{1}{12} \func{\Delta^0}{w} + \ldots}, \notag
\end{align}
the rest being obtained by these from \eqnref{eqnOPECommk=4}.

There is a conformal embedding of $\func{\affine{\alg{sl}}}{2}_4$ into $\func{\affine{\alg{sl}}}{3}_1$, meaning that the corresponding coset model $\func{\affine{\alg{sl}}}{3}_1 / \func{\affine{\alg{sl}}}{2}_4$ has vanishing central charge.  This does not directly imply the equivalence of the two \cfts{}.  Indeed, the spectrum of the $\func{\affine{\alg{sl}}}{3}_1$-\WZW{} theory indicates that the correct $\func{\affine{\alg{sl}}}{2}_4$-\WZW{} theory is the non-diagonal one \cite{BaiAcc86}.  Nevertheless, the chiral algebras of the two theories need to correspond in some way.  We will detail this correspondence by proving that the chiral algebras $\alg{A}_4$ and $\uealg{\func{\affine{\alg{sl}}}{3}_1}$ are isomorphic.

Roughly speaking, the three $\func{\alg{sl}}{2}$ generators are augmented by the five components of the simple current (whose conformal dimension is $1$) to fill out the eight $\func{\alg{sl}}{3}$ generators:
\begin{equation}
\begin{matrix}
& E & & F & \\
\phi^{\brac{0}} & & H , \phi^{\brac{2}} & & \phi^{\brac{4}} \\
& \phi^{\brac{1}} & & \phi^{\brac{3}} &
\end{matrix}
\qquad \longleftrightarrow \qquad
\begin{matrix}
& e_1 & & f_1 & \\
e_{\theta} & & h_1 , h_2 & & f_{\theta} \\
& e_2 & & f_2 &
\end{matrix}
.
\end{equation}

We choose the $\func{\alg{sl}}{3}$ generators in the following way:  For the two simple roots, we let $h_1$ and $h_2$ be the corresponding coroots, and let $e_1$, $f_1$ and $e_2$, $f_2$ be the corresponding root vectors, normalised by $\comm{e_i}{f_i} = h_i$.  The Killing form\footnote{We denote the Killing form of $\func{\alg{sl}}{3}$ by $\otherkilling{\cdot}{\cdot}$ to distinguish it from that of $\func{\alg{sl}}{2}$.} therefore satisfies
\begin{equation}
\otherkilling{e_i}{f_j} = \otherkilling{f_i}{e_j} = \delta_{ij}, \qquad \otherkilling{h_i}{h_j} = A_{ij} \quad \text{($A = 
\begin{pmatrix}
2 & -1 \\ -1 & 2
\end{pmatrix}
$, the Cartan matrix),}
\end{equation}
and vanishes for all other combinations of $e_i$, $f_i$ and $h_i$.  We define $e_{\theta} = \comm{e_1}{e_2}$ and $f_{\theta} = e_{\theta}^{\dag} = \comm{f_2}{f_1}$.  Invariance of the Killing form and the Jacobi identity then give $\otherkilling{e_{\theta}}{f_{\theta}} = - \otherkilling{h_1}{h_2} = 1$, hence
\begin{equation}
\comm{e_2}{f_{\theta}} = f_1, \quad \comm{f_{\theta}}{e_1} = f_2, \quad \comm{e_{\theta}}{f_2} = e_1, \quad \comm{f_1}{e_{\theta}} = e_2
\end{equation}
(invariance), and finally, $\comm{e_{\theta}}{f_{\theta}} = h_1 + h_2$ (Jacobi identity).

The isomorphism between $\alg{A}_4$ and $\uealg{\func{\affine{\alg{sl}}}{3}_1}$ may be derived by comparing the \opes{} of each theory\footnote{Alternatively, we can directly compare the $\func{\affine{\alg{sl}}}{3}_1$ commutation relations with the supercommutation relations
\begin{equation}
\phi^{\brac{m}}_r \phi^{\brac{n}}_s - \brac{-1}^{m+n} \phi^{\brac{n}}_s \phi^{\brac{m}}_r \gcreq{2} \mathcal{S}_{m,n} \sqbrac{J^{\pair{m}{n}}_{r+s} + r \delta_{m+n,4} \delta_{r+s,0}},
\end{equation}
where $J^{\pair{m}{n}}$ denotes $E$, $\tfrac{1}{8} \brac{n-m} H$ or $F$ according as to whether $m+n$ is $3$, $4$ or $5$ (respectively), and vanishes otherwise.}.  If we suppose that $H = a h_1 + b h_2$ and that $\phi^{\brac{0}}$ is proportional to $e_{\theta}$, then by comparing
\begin{equation}
\radord{\func{H}{z} \func{\phi^{\brac{0}}}{w}} = \frac{4 \func{\phi^{\brac{0}}}{w}}{z-w} + \ldots \quad \text{and} \quad \radord{\func{\brac{a h_1 + b h_2}}{z} \func{e_{\theta}}{w}} = \frac{\brac{a+b} \func{e_{\theta}}{w}}{z-w} + \ldots,
\end{equation}
we derive that $a+b = 4$.  Similarly, comparing $\radord{\func{H}{z} \func{H}{w}}$ with its $\func{\affine{\alg{sl}}}{3}_1$ counterpart gives $a^2 - ab + b^2 = 4$, hence $a = b = 2$.  If we now put $\phi^{\brac{2}} = \alpha h_1 + \beta h_2$, then considering its \opes{} with $H$ and with itself, we conclude that $\beta = -\alpha$ and $\alpha^2 = \mathcal{S}_{2,2} / 6$.

At this point we pause to note that the $\func{\affine{\alg{sl}}}{3}$ currents are all mutually bosonic, whereas our simple current fields are not (\eqnref{eqnOPECommk=4}).  Again, we must remedy this with an operator $\mathcal{F}$, and \eqnref{eqnFDef} advises that we must put
\begin{equation}
\phi^{\brac{2}} \mathcal{F} = \sqrt{\frac{\mathcal{S}_{2,2}}{6}} \brac{h_1 - h_2}.
\end{equation}
The remaining calculations are routine and offer no new subtleties.  The final result is as follows:
\begin{equation} \label{Isok=4}
\begin{array}{>{\displaystyle}c}
\phi^{\brac{0}} \mathcal{F} = - \sqrt{\mathcal{S}_{2,2}} e_{\theta} \vphantom{e_{\theta} = \frac{\mathcal{F}}{\sqrt{\mathcal{S}_{2,2}}} \phi^{\brac{0}}} \\
E = \sqrt{2} \brac{e_1 + e_2} \vphantom{e_1 = \frac{1}{2 \sqrt{2}} E + \frac{\mathcal{F}}{\sqrt{2 \mathcal{S}_{2,2}}} \phi^{\brac{1}}} \\
\phi^{\brac{1}} \mathcal{F} = \sqrt{\frac{\mathcal{S}_{2,2}}{2}} \brac{e_1 - e_2} \vphantom{e_2 = \frac{1}{2 \sqrt{2}} E - \frac{\mathcal{F}}{\sqrt{2 \mathcal{S}_{2,2}}} \phi^{\brac{1}}} \\
H = 2 \brac{h_1 + h_2} \vphantom{h_1 = \frac{1}{4} H + \mathcal{F} \sqrt{\frac{3}{2 \mathcal{S}_{2,2}}} \phi^{\brac{2}}} \\
\phi^{\brac{2}} \mathcal{F} = \sqrt{\frac{\mathcal{S}_{2,2}}{6}} \brac{h_1 + h_2} \vphantom{h_2 = \frac{1}{4} H - \mathcal{F} \sqrt{\frac{3}{2 \mathcal{S}_{2,2}}} \phi^{\brac{2}}} \\
F = \sqrt{2} \brac{f_1 + f_2} \vphantom{f_1 = \frac{1}{2 \sqrt{2}} F - \frac{\mathcal{F}}{\sqrt{2 \mathcal{S}_{2,2}}} \phi^{\brac{3}}} \\
\phi^{\brac{3}} \mathcal{F} = - \sqrt{\frac{\mathcal{S}_{2,2}}{2}} \brac{f_1 - f_2} \vphantom{f_2 = \frac{1}{2 \sqrt{2}} F + \frac{\mathcal{F}}{\sqrt{2 \mathcal{S}_{2,2}}} \phi^{\brac{3}}} \\
\phi^{\brac{4}} \mathcal{F} = -\sqrt{\mathcal{S}_{2,2}} f_{\theta} \vphantom{f_{\theta} = \frac{\mathcal{F}}{\sqrt{\mathcal{S}_{2,2}}} \phi^{\brac{4}}}
\end{array}
\qquad \Longleftrightarrow \qquad
\begin{array}{>{\displaystyle}c}
\vphantom{\phi^{\brac{0}} = \sqrt{\mathcal{S}_{2,2}} e_{\theta}} e_{\theta} = \frac{-1}{\sqrt{\mathcal{S}_{2,2}}} \phi^{\brac{0}} \mathcal{F} \\
\vphantom{E = \sqrt{2} \brac{e_1 + e_2} \vphantom{e_1 = \frac{1}{2 \sqrt{2}} E + \frac{\mathcal{F}}{\sqrt{2 \mathcal{S}_{2,2}}} \phi^{\brac{1}}}} e_1 = \frac{1}{2 \sqrt{2}} E + \frac{1}{\sqrt{2 \mathcal{S}_{2,2}}} \phi^{\brac{1}} \mathcal{F} \\
\vphantom{\phi^{\brac{1}} = \mathcal{F} \sqrt{\frac{\mathcal{S}_{2,2}}{2}} \brac{e_1 - e_2}} e_2 = \frac{1}{2 \sqrt{2}} E - \frac{1}{\sqrt{2 \mathcal{S}_{2,2}}} \phi^{\brac{1}} \mathcal{F} \\
\vphantom{H = 2 \brac{h_1 + h_2}} h_1 = \frac{1}{4} H + \sqrt{\frac{3}{2 \mathcal{S}_{2,2}}} \phi^{\brac{2}} \mathcal{F} \\
\vphantom{\phi^{\brac{2}} = \mathcal{F} \sqrt{\frac{\mathcal{S}_{2,2}}{6}} \brac{h_1 + h_2}} h_2 = \frac{1}{4} H - \sqrt{\frac{3}{2 \mathcal{S}_{2,2}}} \phi^{\brac{2}} \mathcal{F} \\
\vphantom{F = \sqrt{2} \brac{f_1 + f_2}} f_1 = \frac{1}{2 \sqrt{2}} F - \frac{1}{\sqrt{2 \mathcal{S}_{2,2}}} \phi^{\brac{3}} \mathcal{F} \\
\vphantom{\phi^{\brac{3}} = - \mathcal{F} \sqrt{\frac{\mathcal{S}_{2,2}}{2}} \brac{f_1 - f_2}} f_2 = \frac{1}{2 \sqrt{2}} F + \frac{1}{\sqrt{2 \mathcal{S}_{2,2}}} \phi^{\brac{3}} \mathcal{F} \\
\vphantom{\phi^{\brac{4}} = \mathcal{F} \sqrt{\mathcal{S}_{2,2}} f_{\theta}} f_{\theta} = \frac{-1}{\sqrt{\mathcal{S}_{2,2}}} \phi^{\brac{4}} \mathcal{F}
\end{array}
.
\end{equation}
We note that the adjoints of $\alg{A}_4$ and $\func{\affine{\alg{sl}}}{3}$ are also preserved, hence we have another isomorphism of $^*$-algebras.  We mention that if we had tried to continue these computations without such an $\mathcal{F}$, we would have quickly reached a contradiction in the algebra.

It is not difficult to verify that this isomorphism identifies the Virasoro fields of our two theories.  Of more interest is the fact that $H = 2 \brac{h_1 + h_2}$ limits the allowed $\alg{A}_4$-\hwss{} to $\ket{0}$ and $\ket{\psi_2}$, as the eigenvalues of $H$ are constrained to be even (this is consistent with the formulae given for $h_1$ and $h_2$ if we remember that $\mathcal{S}_{2,2}$ has eigenvalue $6$ on $\ket{\psi_2}$).  Actually, this is just the constraint imposed by monodromy charge:  Linear combinations such as those appearing on the right of (\ref{Isok=4}) only make sense at the level of modes when acting upon states whose charge is integral, so that the mode subscripts are integers.

Assuming that $\phi^{\brac{2}}_0$ interchanges the \hwss{} $\ket{\psi_2}$ and $\ket{\widetilde{\psi}_2}$, the precise correspondence is then
\begin{equation}
\ket{0} \quad \longleftrightarrow \quad \ket{\pair{0}{0}} \qquad \text{and} \qquad 
\begin{matrix}
\ket{\psi_2} + \ket{\widetilde{\psi}_2} \vphantom{\ket{\pair{1}{0}}} \\
\ket{\psi_2} - \ket{\widetilde{\psi}_2} \vphantom{\ket{\pair{0}{1}}}
\end{matrix}
\quad \longleftrightarrow \quad
\begin{matrix}
\vphantom{\ket{\psi_2} + \ket{\widetilde{\psi}_2}} \ket{\pair{1}{0}} \\
\vphantom{\ket{\psi_2} - \ket{\widetilde{\psi}_2}} \ket{\pair{0}{1}}
\end{matrix}
,
\end{equation}
where $\ket{\pair{\lambda_1}{\lambda_2}}$ denotes the $\func{\affine{\alg{sl}}}{3}_1$-\hwss{} whose $h_i$-eigenvalue is $\lambda_i$.  This correspondence between highest weight states can be lifted to the irreducible modules and their characters.  To see this, we now turn to a study of the singular vectors of the $\alg{A}_k$-Verma modules.

\section{Singular Vectors} \label{secSingVects}

\subsection{Preliminary Remarks} \label{secSingVectsPrelim}

Recall from \secref{secRepHWMs} that we have identified two affine \hwss{} in each $\alg{A}_k$-Verma module:  The $\alg{A}_k$-\hws{} itself, and its $\alg{A}_k$-descendant given in \eqnref{eqnIdentifyAffHWSs}.  That there are only two (non-singular) affine \hwss{} is to be expected as the simple current has order two in the fusion ring.  Nevertheless, we can give a direct proof of this fact using the following lemma, which states that the action of any bilinear in the $\phi$-modes within a \hwm{} can be replaced by a combination of affine modes.

\begin{lemma} \label{lemTwoPhi=Aff}
Any state of the form $\phi^{\brac{m}}_r \phi^{\brac{n}}_s \ket{\psi}$, where $\ket{\psi}$ is an arbitrary state in a \hwm{}, may be written as a linear combination of affine descendants of $\ket{\psi}$.
\end{lemma}
\begin{proof}
The \gcrs{} given in \eqnref{eqnGeneralGCRs}, for difference choices of $\gamma$, all express an infinite linear combination of quadratics in the $\phi$-modes, in terms of pure affine modes (recall that the $\mathcal{S}_{m,n}$ act as constants within \hwms{}).  We therefore seek to invert this relation.  Write the \gcrs{} in the form
\begin{equation}
\sum_{\ell = 0}^{\infty} \binom{\ell - k/2 + \gamma - 1}{\ell} \sqbrac{\phi^{\brac{m}}_{r - \ell} \phi^{\brac{n}}_{s + \ell} - \brac{-1}^{k-m-n + \gamma} \phi^{\brac{n}}_{s + k/2 - \gamma - \ell} \phi^{\brac{m}}_{r - k/2 + \gamma + \ell}} \ket{\psi} = \func{R^{\pair{m}{n}}_{r,s}}{\gamma} \ket{\psi},
\end{equation}
where $\func{R^{\pair{m}{n}}_{r,s}}{\gamma}$ denotes the right-hand-side of \eqnref{eqnGeneralGCRs}.  We determine an integer $N \geqslant 0$ such that $\phi^{\brac{n}}_{s + N + \ell}$ annihilates $\ket{\psi}$ for all $\ell \geqslant 0$, and then choose $\gamma$ sufficiently large so that $\phi^{\brac{m}}_{r - N - k/2 + \gamma + \ell}$ also annihilates $\ket{\psi}$ for all $\ell \geqslant 0$.  The \gcrs{} now become (with $r \rightarrow r-i$ and $s \rightarrow s+i$)
\begin{equation}
\sum_{\ell = 0}^{N-1} \binom{\ell - k/2 + \gamma - 1}{\ell} \phi^{\brac{m}}_{r - i - \ell} \phi^{\brac{n}}_{s + i + \ell} \ket{\psi} = \func{R^{\pair{m}{n}}_{r-i,s+i}}{\gamma} \ket{\psi} \qquad \text{(for $i=0, \ldots, N-1$).}
\end{equation}
This is a set of linear equations for the $\phi^{\brac{m}}_{r-i - \ell} \phi^{\brac{n}}_{s+i + \ell} \ket{\psi}$, and the corresponding matrix is triangular with diagonal entries equal to one, hence invertible.  It therefore follows that each $\phi^{\brac{m}}_{r-i - \ell} \phi^{\brac{n}}_{s+i + \ell} \ket{\psi}$ can be expressed as a linear combination of the $\func{R^{\pair{m}{n}}_{r-j,s+j}}{\gamma} \ket{\psi}$.  But these are expressed as affine modes acting on $\ket{\psi}$, as required.
\end{proof}

We are now in a position to prove the announced result, namely:

\begin{proposition} \label{propTwoAffHWSs}
There are only two non-singular affine \hwss{} in each $\alg{A}_k$-Verma module (though they may coincide).
\end{proposition}
\begin{proof}
We have already identified two such \hwss{}, $\ket{\psi_{\lambda}}$ and $\ket{\psi_{k - \lambda}} = \phi^{\brac{\lambda}}_{\lambda / 2 - k/4} \ket{\psi_{\lambda}}$, in each $\alg{A}_k$-Verma module.  We therefore need only show that there are no others.

Consider therefore an arbitrary state of the form $\phi^{\brac{m_1}}_{r_1} \cdots \phi^{\brac{m_{\ell}}}_{r_{\ell}} \ket{\psi_{\lambda}}$.  If $\ell = 0$, then we are considering the $\alg{A}_k$-\hws{}.  If $\ell = 1$, then we are considering $\phi^{\brac{m}}_r \ket{\psi_{\lambda}}$, and \eqnref{eqnFirstDescendant} gives $r \leqslant \theta^{\pair{m}{0}}_{\lambda} - k/4$, where $\theta^{\pair{m}{0}}_{\lambda}$ is $\lambda / 2$ or $m - \lambda / 2$ according as to whether $m \geqslant \lambda$ or not.  Acting on these states with $E_0$ and $F_1$ suffices to show that we have already found the only affine \hws{} with $\ell = 1$.  It follows that every other state with $\ell = 1$ must be an affine descendant of this \hws{}, as an affine descendant of $\ket{\psi_{\lambda}}$ must be expressed in terms of at least two $\phi$-modes (as follows from the \gcrs{}).

But if $\ell > 1$, \lemref{lemTwoPhi=Aff} allows us to replace the first two $\phi$-modes with (a linear combination of) affine modes.  Thus, $\phi^{\brac{m_1}}_{r_1} \cdots \phi^{\brac{m_{\ell}}}_{r_{\ell}} \ket{\psi_{\lambda}}$ is an affine descendant of $\phi^{\brac{m_3}}_{r_3} \cdots \phi^{\brac{m_{\ell}}}_{r_{\ell}} \ket{\psi_{\lambda}}$.  Inductively, it follows that states with $\ell > 1$ are affine descendants of $\ket{\psi_{\lambda}}$ or $\ket{\psi_{k - \lambda}}$, according as to whether $\ell$ is even or odd (respectively).  Any such descendants which are also affine \hwss{} are singular, hence the proposition is proved.
\end{proof}

Let $\mathcal{M}^{\brac{k}}_{\lambda}$ denote the $\alg{A}_k$-Verma module of highest weight $\lambda$, and let $\affine{\mathcal{M}}^{\brac{k}}_{\lambda}$ denote the corresponding $\func{\affine{\alg{sl}}}{2}_k$-Verma module.  Then, \propref{propTwoAffHWSs} implies that the inclusion of $\uealg{\func{\affine{\alg{sl}}}{2}_k}$ into $\alg{A}_k$ induces a \emph{surjection} of the form
\begin{equation} \label{eqnSurjection}
\affine{\mathcal{M}}^{\brac{k}}_{\lambda} \oplus \affine{\mathcal{M}}^{\brac{k}}_{k - \lambda} \longrightarrow \mathcal{M}^{\brac{k}}_{\lambda}.
\end{equation}
Furthermore, our simultaneous normalisation of all zero-grade states (in particular of the affine \hwss) implies that this surjection is an isometry.  It follows that the singular vectors of the affine modules (affine singular vectors for short) must therefore be mapped to $\alg{A}_k$-singular vectors, and \emph{all} $\alg{A}_k$-singular vectors arise in this way.

We can now investigate the $\alg{A}_k$-singular vectors by determining the kernel of our surjection (\ref{eqnSurjection}).  What we will find is that this kernel coincides with the set of affine singular vectors.  In other words, each affine singular vector is mapped to zero in the $\alg{A}_k$-Verma module, so the $\alg{A}_k$-Verma module is actually irreducible, and decomposes as the direct sum of two \emph{irreducible} affine modules (this is the content of \corref{corModuleIso}).  Before demonstrating this result, we verify it in the following section within the context of the two simplest models, $k = 1$ and $2$.

\subsection{Some Examples} \label{secSingVectsEx}

The easiest case to analyse is of course $k=1$.  The \opes{} defining the extended algebra are as follows:
\begin{align}
\radord{\func{\phi^{\brac{0}}}{z} \func{\phi^{\brac{0}}}{w}} &= \func{E}{w} \brac{z-w}^{1/2} + \tfrac{1}{2} \func{\partial E}{w} \brac{z-w}^{3/2} + \ldots \notag \\
\radord{\func{\phi^{\brac{0}}}{z} \func{\phi^{\brac{1}}}{w}} &= \frac{1}{\brac{z-w}^{1/2}} + \tfrac{1}{2} \func{H}{w} \brac{z-w}^{1/2} + \sqbrac{\tfrac{1}{2} \func{T}{w} + \tfrac{1}{4} \func{\partial H}{w}} \brac{z-w}^{3/2} + \ldots \\
\radord{\func{\phi^{\brac{1}}}{z} \func{\phi^{\brac{1}}}{w}} &= \func{F}{w} \brac{z-w}^{1/2} + \tfrac{1}{2} \func{\partial F}{w} \brac{z-w}^{3/2} + \ldots \notag
\end{align}
The $\gamma = 2$ \gcr{} is therefore
\begin{equation}
\sum_{\ell = 0}^{\infty} \binom{\ell + \tfrac{1}{2}}{\ell} \sqbrac{\phi^{\brac{m}}_{r-\ell} \phi^{\brac{n}}_{s+\ell} + \brac{-1}^{m+n} \phi^{\brac{n}}_{s-3/2-\ell} \phi^{\brac{m}}_{r+3/2+\ell}} \gcreq{2} J^{\pair{m}{n}}_{r+s},
\end{equation}
where $J^{\pair{m}{n}}$ stands for $E$, $\tfrac{1}{2} \brac{n-m} H$, or $F$, according as whether $m+n$ is $0$, $1$, or $2$, respectively (note also that $\mathcal{S}_{m,n} = \id$ for all $m$ and $n$).

The singular vectors of the affine modules are generated by $F_0 \ket{0}$, $F_0 \ket{\phi^{\brac{1}}}$, $E_{-1} \ket{\phi^{\brac{0}}}$, and $E_{-1}^2 \ket{0}$.  But, if we take $m=n=1$ and $s=3/4$, we find that
\begin{equation}
F_0 \ket{0} \gcreq{2} \sum_{\ell = 0}^{\infty} \binom{\ell + \tfrac{1}{2}}{\ell} \sqbrac{\phi^{\brac{1}}_{-3/4-\ell} \phi^{\brac{1}}_{3/4+\ell} + \phi^{\brac{1}}_{-3/4-\ell} \phi^{\brac{1}}_{3/4+\ell}} \ket{0} = 0.
\end{equation}
as $\phi^{\brac{1}}_r$ annihilates the vacuum for all $r > \tfrac{-1}{4}$.  Similarly, we find that $F_0 \ket{\phi^{\brac{1}}}$ vanishes when we take $m=n=1$ and $s=3/4$, because $\phi^{\brac{1}}_r$ annihilates $\ket{\phi^{\brac{1}}}$ when $r > \theta^{\pair{1}{1}}_1 - \tfrac{1}{4} = \tfrac{-3}{4}$ (\secref{secRepCharges}).  Putting $m=n=0$ and $s=\tfrac{1}{4}$, we learn that $E_{-1} \ket{\phi^{\brac{0}}}$ also vanishes identically (we use $\theta^{\pair{0}{0}}_1 = \tfrac{1}{2}$), so we are left with $E_{-1}^2 \ket{0}$.  The demonstration of vanishing is less immediate in this case, and may be discerned by noting that
\begin{equation}
E_{-1}^2 \ket{0} \gcreq{2} E_{-1} \phi^{\brac{0}}_{-3/4} \phi^{\brac{0}}_{-1/4} \ket{0} = - \phi^{\brac{0}}_{-3/4} E_{-1} \ket{\phi^{\brac{0}}} = 0,
\end{equation}
since we have already established that $E_{-1} \ket{\phi^{\brac{0}}}$ vanishes.

When $k=2$, we have three affine modules, hence six primitive singular vectors to consider.  The $\gamma = 1$ \gcr{} was given in \eqnref{eqnSuperCommk=2}, and $\gamma = 2$ gives
\begin{equation}
\sum_{\ell = 0}^{\infty} \sqbrac{\phi^{\brac{m}}_{r-\ell} \phi^{\brac{n}}_{s+\ell} - \brac{-1}^{m+n} \phi^{\brac{n}}_{s-1-\ell} \phi^{\brac{m}}_{r+1+\ell}} \gcreq{2} \mathcal{S}_{m,n} J^{\pair{m}{n}}_{r+s},
\end{equation}
where $J^{\pair{m}{n}}$ now stands for $E$, $\tfrac{1}{4} \brac{n-m} H$, or $F$, according as whether $m+n$ is $1$, $2$, or $3$, respectively (and vanishes if $m+n$ is $0$ or $4$).

As with the $k=1$ case, the singular vectors $F_0 \ket{0}$, $F_0 \ket{\psi_1^{\brac{1}}}$, and $F_0 \ket{\phi^{\brac{2}}}$ can be easily checked to vanish identically when the $\gamma = 2$ \gcr{} is applied with $n=2$ and $s = \tfrac{1}{2}$, $0$, or $\tfrac{1}{2}$, respectively.  Similarly, $E_{-1} \ket{\phi^{\brac{0}}}$ vanishes identically upon taking $s = \tfrac{-1}{2}$ and $n = 0$.  This then gives
\begin{equation}
E_{-1}^3 \ket{0} \gcreq{2} \sqrt{2} E_{-1}^2 \phi^{\brac{1}}_{-1/2} \ket{\phi^{\brac{0}}} = 2 E_{-1} \phi^{\brac{0}}_{-3/2} \ket{\phi^{\brac{0}}} = 0,
\end{equation}
where we anticommute the $E_{-1}$ to the right, where they annihilate $\ket{\phi^{\brac{0}}}$.

Verifying the vanishing of the last singular vector requires a little more delicacy.  One straight-forward demonstration is as follows:
\begin{align}
E_{-1}^2 \ket{\psi_1^{\brac{0}}} &\gcreq{2} \frac{1}{\sqrt{2}} E_{-1} \phi^{\brac{0}}_{-1} \phi^{\brac{1}}_0 \ket{\psi_1^{\brac{0}}} = \frac{-1}{\sqrt{2}} \phi^{\brac{0}}_{-1} E_{-1} \ket{\widetilde{\psi}_1^{\brac{0}}} \gcreq{2} \frac{-1}{2} \phi^{\brac{0}}_{-1} \phi^{\brac{0}}_{-1} \phi^{\brac{1}}_0 \phi^{\brac{1}}_0 \ket{\psi_1^{\brac{0}}} \notag \\
&\gcreq{1} \frac{-1}{4} \phi^{\brac{0}}_{-1} \phi^{\brac{0}}_{-1} \mathcal{S}_{1,1} \ket{\psi_1^{\brac{0}}} = \frac{-1}{2} \phi^{\brac{0}}_{-1} \phi^{\brac{0}}_{-1} \ket{\psi_1^{\brac{0}}} \gcreq{1} 0.
\end{align}
A more elegant demonstration involves using the $\gamma = 3$ \gcr{},
\begin{equation}
\sum_{\ell = 0}^{\infty} \brac{\ell + 1} \sqbrac{\phi^{\brac{0}}_{r-\ell} \phi^{\brac{0}}_{s+\ell} + \phi^{\brac{0}}_{s-2-\ell} \phi^{\brac{0}}_{r+2+\ell}} \gcreq{3} \mathcal{S}_{0,0} \sum_{t \in \ZZ} \normord{E_t E_{r+s-t}},
\end{equation}
and noting that $\sum_t \normord{E_t E_{-2-t}} \ket{\psi} = E_{-1}^2 \ket{\psi}$ when $\ket{\psi}$ is an affine \hws{}:
\begin{equation}
E_{-1}^2 \ket{\psi_1^{\brac{0}}} \gcreq{3} \phi^{\brac{0}}_{-1} \phi^{\brac{0}}_{-1} \ket{\psi_1^{\brac{0}}} \gcreq{1} 0.
\end{equation}
This belies the fact that the vanishing may be demonstrated using \gcrs{} with $\gamma < 3$.  Nevertheless, it does suggest a strategy by which this vanishing may be demonstrated for general $k$.

\subsection{Vanishing in General} \label{secSingVectsGenVan}

Remarkably, it is possible to demonstrate explicitly that all affine singular vectors vanish identically when the algebra is extended, hence that these extended algebras define faithful (or free-field-like) representations of the $\func{\group{SU}}{2}$ \WZW{} models.  This may sound surprising, but recall from \secref{secAlgOPEs} that we have used the structure of the \emph{irreducible} affine vacuum module to derive the extended algebra, and in \secref{secRepCharges} we used the structures of the other \emph{irreducible} modules to derive the $\func{\alg{u}}{1}$-valued charges $\theta^{\pair{m}{n}}_{\lambda}$.  A knowledge of the affine singular vectors was therefore built in to the extension, in a sense, so it should not be surprising to find that they do not appear in the extended theory.  This suggests that the extended algebra approach we have constructed above should be interpreted as a \emph{natural} ``free-field'' reformulation of the $\func{\group{SU}}{2}$ \WZW{} models.

\begin{theorem} \label{thmSingVectsVanish}
The affine singular vectors generated by $F_0^{\lambda + 1} \ket{\psi_{\lambda}}$ and $E_{-1}^{k - \lambda + 1} \ket{\psi_{\lambda}}$, for $\lambda = 0, \ldots, k$, span the kernels of the surjections of \eqnref{eqnSurjection}.
\end{theorem}
\begin{proof}
It is enough to show that the primitive singular vectors vanish in the corresponding $\alg{A}_k$-modules.  We consider then $F_0^{\lambda + 1} \ket{\psi_{\lambda}}$.  As in the proof of \propref{propSEigs}, we note that
\begin{equation}
\sum_{\substack{t_1, \ldots, t_j \in \ZZ \\ t_1 + \ldots + t_j = 0}} \normord{F_{t_1} \cdots F_{t_j}} \ket{\psi} = F_0^j \ket{\psi}
\end{equation}
when $\ket{\psi}$ is a \hws{}, and that this is just (a multiple of) the right-hand-side of the \gcr{} with $\gamma = j+1$, $m+n = j+k$ and $r+s = 0$, applied to $\ket{\psi}$.

We therefore take $\gamma = \lambda + 2$, $m+n = k + \lambda + 1$, and choose $-r = s = \lambda / 2 - k/4 + 1$.  This gives
\begin{multline}
F_0^{\lambda + 1} \ket{\psi_{\lambda}} \gcreq{\lambda + 2} \mathcal{S}_{m,n}^{-1} \sum_{\ell = 0}^{\infty} \binom{\ell - k/2 + \lambda + 1}{\ell} \\*
\cdot \sqbrac{\phi^{\brac{m}}_{k/4 - \lambda / 2 - 1 - \ell} \phi^{\brac{n}}_{\lambda / 2 + 1 - k/4 + \ell} + \phi^{\brac{n}}_{k/4 - \lambda / 2 - 1 - \ell} \phi^{\brac{m}}_{\lambda / 2 + 1 - k/4 + \ell}} \ket{\psi_{\lambda}} = 0,
\end{multline}
because $\theta^{\pair{m}{0}}_{\lambda} \leqslant \lambda / 2$ for all $m$.  This proves that this singular vector vanishes in all cases when $m$ and $n$ can be chosen to sum to $k + \lambda + 1$.  This covers all cases except $\lambda = k$, for which we offer the simple argument that
\begin{equation}
F_0^{k+1} \ket{\phi} = F_0^{k+1} \phi^{\brac{0}}_{-k/4} \ket{0} = 0,
\end{equation}
as $F_0$ has already been shown to annihilate $\ket{0}$, so anticommuting the $F_0$ to the right eventually gives something proportional to $F_0 \phi^{\brac{k}}_{-k/4} \ket{0}$, which obviously vanishes upon one further anticommutation.

The demonstration for $E_{-1}^{k - \lambda + 1} \ket{\psi_{\lambda}}$ is similar if more intricate.  The fact that $\ket{\psi_{\lambda}}$ is an affine \hws{} again leads to
\begin{equation}
E_{-1}^{k - \lambda + 1} \ket{\psi_{\lambda}} = \sum_{\substack{t_1, \ldots, t_{k - \lambda + 1} \in \ZZ \\ t_1 + \ldots + t_{k - \lambda + 1} = -k + \lambda - 1}} \normord{E_{t_1} \cdots E_{t_{k - \lambda + 1}}} \ket{\psi_{\lambda}},
\end{equation}
which is (a multiple of) the right-hand-side of the \gcr{} with $\gamma = k - \lambda + 2$ and $m+n = \lambda - 1$, applied to $\ket{\psi_{\lambda}}$.  The conformal dimension determines $r+s$ to be $-k + \lambda - 1$.

When $\lambda$ is odd, we choose $s = 1/2 - k/4$, so the modes acting immediately on $\ket{\psi_{\lambda}}$ are $\phi^{\brac{n}}_{1/2 - k/4 + \ell}$ and $\phi^{\brac{m}}_{1/2 - k/4 + \ell}$.  Since $m+n = \lambda - 1$, we have $\theta^{\pair{m}{0}}_{\lambda} = m - \lambda / 2$ and $\theta^{\pair{n}{0}}_{\lambda} = n - \lambda / 2$, hence these modes will annihilate $\ket{\psi_{\lambda}}$ if $m$ and $n$ are both less than $\brac{\lambda + 1} / 2$.  We therefore choose $m = n = \brac{\lambda - 1} / 2$, demonstrating that this singular vector vanishes identically.  When $\lambda$ is even, we cannot choose $s = 1/2 - k/4$ due to the monodromy charge of $\ket{\psi_{\lambda}}$.  Instead, we take $s = -k/4$, so the modes acting on $\ket{\psi_{\lambda}}$ are $\phi^{\brac{n}}_{-k/4 + \ell}$ and $\phi^{\brac{m}}_{1 - k/4 + \ell}$.  Complete annihilation will then occur if $n < \lambda / 2$ and $m < \lambda / 2 + 1$.  Taking $m = \lambda / 2$ and $n = \lambda / 2 - 1$ then proves that the singular vector also vanishes identically in this case.

Finally, we again note that $m+n = \lambda - 1$ cannot be satisfied when $\lambda = 0$, but we can quickly verify that
\begin{equation}
E_{-1}^{k + 1} \ket{0} \gcreq{2} \mathcal{S}_{k-1,0}^{-1} E_{-1}^k \phi^{\brac{k-1}}_{k/4 - 1} \phi^{\brac{0}}_{-k/4} \ket{0} = \mathcal{S}_{k-1,0}^{-1} E_{-1}^k \phi^{\brac{k-1}}_{k/4 - 1} \ket{\phi} = 0,
\end{equation}
by anticommuting the $E_{-1}$ to the right where they annihilate $\ket{\phi}$.
\end{proof}

\begin{corollary} \label{corModuleIso}
The $\alg{A}_k$-Verma module is irreducible, and is isomorphic to the direct sum of two affine irreducible \hwms{} (irreducible modules are denoted by $\mathcal{L}$):
\begin{equation}
\affine{\mathcal{L}}^{\brac{k}}_{\lambda} \oplus \affine{\mathcal{L}}^{\brac{k}}_{k - \lambda} \cong \mathcal{M}^{\brac{k}}_{\lambda} = \mathcal{L}^{\brac{k}}_{\lambda}.
\end{equation}
\end{corollary}

\subsection{Consequences when $k = 2$ or $4$} \label{secExamplesRevisited}

To complete our study of singular vectors, let us consider what the vanishing of the singular vectors implies for the examples studied in \secref{secExamples}.  In \secref{secExamplek=2}, we showed that $\alg{A}_2$ was isomorphic to the tensor product of three copies of the extended algebra, $\overline{\alg{A}}$, of the Ising model (that is, three free fermions $\chi^+$, $\chi^0$ and $\chi^-$).  The $\alg{A}_2$-vacuum thereby decomposed into the tensor product of three fermionic vacuum states, and the $\alg{A}_2$-\hws{} $\ket{\psi_1}$ was identified with the product of three copies of the fermionic \hws{} of conformal dimension $\tfrac{1}{16}$.  The vanishing of the singular vectors, both in the $\alg{A}_2$-vacuum module and the fermionic vacuum modules, then leads to the conclusion that we have an isomorphism of modules (graded by conformal dimension only):
\begin{equation}
\affine{\mathcal{L}}^{\brac{2}}_{0} \oplus \affine{\mathcal{L}}^{\brac{2}}_{2} \cong \mathcal{L}^{\brac{2}}_0 \cong \overline{\mathcal{L}}_0 \otimes \overline{\mathcal{L}}_0 \otimes \overline{\mathcal{L}}_0,
\end{equation}
where $\overline{\mathcal{L}}_h$ denotes the fermionic module whose \hws{} has conformal dimension $h$.  We shall explicitly confirm this fact in \secref{secChark=2}, by showing that the character of the $\alg{A}_2$-vacuum module $\mathcal{L}^{\brac{2}}_0$ is the cube of the character of the fermionic vacuum module $\overline{\mathcal{L}}_{0}$.

Na\"{\i}vely, we might suppose that the character of $\mathcal{L}^{\brac{2}}_1$ is also the perfect cube of the character of $\overline{\mathcal{L}}_{1/16}$.  However, this is not correct (see \secref{secChark=2}), though it is nearly so.  The obstruction is subtle:  The chiral algebras are $^*$-isomorphic and there are no singular vectors to consider, in $\mathcal{L}^{\brac{2}}_1$ by \thmref{thmSingVectsVanish} and in $\overline{\mathcal{L}}_{1/16}$ as is well-known.  But, the natural definition of a \hwm{} is slightly different for the two algebras, and it is this difference which is not preserved by our isomorphism.  Explicitly, $\ket{\psi_1}$ is an affine \hws{}, so it is annihilated by $E_0$, whereas the fermionic \hws{} $\ket{\overline{\frac{1}{16}}}$ is only a Virasoro \hws{} (by definition), hence it is not annihilated by any mode of zero-grade.  In other words, this slight incompatibility in the respective definitions leads to 
\begin{equation}
0 = E_0 \ket{\psi_1} \gcreq{2} \phi^{\brac{1}}_0 \phi^{\brac{0}}_0 \ket{\psi_1} \longleftrightarrow \frac{\ii}{\sqrt{2}} \sqbrac{\chi_0 \ket{\overline{\tfrac{1}{16}}} \otimes \chi_0 \ket{\overline{\tfrac{1}{16}}} \otimes \ket{\overline{\tfrac{1}{16}}} + \ii \ket{\overline{\tfrac{1}{16}}} \otimes \chi_0 \ket{\overline{\tfrac{1}{16}}} \otimes \chi_0 \ket{\overline{\tfrac{1}{16}}}} \neq 0,
\end{equation}
using the explicit form of the isomorphism given in \secref{secExamplek=2}.

It should be clear that this difference is only relevant for zero-grade modes, and that it was not relevant in our consideration of the vacuum module.  Explicitly, we have $8$ independent states at zero-grade in $\overline{\mathcal{L}}_{1/16}^{\otimes 3}$ (each $\ket{\overline{\tfrac{1}{16}}}$ may have a $\chi_0$ acting on it or not), but only $4$ such states in $\mathcal{L}^{\brac{2}}_1$:
\begin{equation}
\ket{\psi_1} = \ket{\psi^{\brac{0}}_1}, \quad \ket{\widetilde{\psi}^{\brac{0}}_1} = \phi^{\brac{1}}_0 \ket{\psi^{\brac{0}}_1}, \quad \ket{\psi^{\brac{1}}_1} = \tfrac{1}{\sqrt{2}} F_0 \ket{\psi^{\brac{0}}_1} \quad \text{and} \quad \ket{\widetilde{\psi}^{\brac{1}}_1} = \tfrac{1}{\sqrt{2}} F_0 \ket{\widetilde{\psi}^{\brac{0}}_1}.
\end{equation}
The isomorphism between the chiral algebras and the absence of singular vectors now allow us to conclude that $\overline{\mathcal{L}}_{1/16}^{\otimes 3}$ has exactly twice as many independent states at each conformal grade as $\mathcal{L}^{\brac{2}}_1$.  In other words, the character of $\mathcal{L}^{\brac{2}}_1$ is precisely \emph{half} the cube of the character of $\overline{\mathcal{L}}_{1/16}$.  We will use this fact in \secref{secChark=2}.

The isomorphism $\alg{A}_4 \cong \uealg{\func{\affine{\alg{sl}}}{3}_1}$ constructed in \secref{secExamplek=4} also extends to isomorphisms between modules.  Because this is a $^*$-isomorphism, norms are preserved, hence \thmref{thmSingVectsVanish} implies that any singular vectors in the $\func{\affine{\alg{sl}}}{3}_1$-modules correspond to $0$ (this can of course be checked explicitly).  The definitions of \hws{} are consistent in this case (we did after all model the $\alg{A}_k$-definition on the affine one).  This is obvious for positively-graded modes, and for the zero-grade modes we note that 
\begin{equation}
E_0 \ket{\psi} = \phi^{\brac{1}}_0 \ket{\psi} = \phi^{\brac{0}}_0 \ket{\psi} = 0 \quad \Longleftrightarrow \quad \brac{e_1}_0 \ket{\psi} = \brac{e_2}_0 \ket{\psi} = \brac{e_{\theta}}_0 \ket{\psi} = 0,
\end{equation}
using the explicit correspondence (\ref{Isok=4}) (and similarly for $H_0$, $\phi^{\brac{2}}_0$, $\brac{h_1}_0$ and $\brac{h_2}_0$).

Denoting the irreducible $\func{\affine{\alg{sl}}}{3}_1$-module whose $\func{\alg{sl}}{3}$-highest weight is $\pair{a}{b}$ by $\affine{\mathcal{V}}^{\brac{1}}_{\pair{a}{b}}$, we have therefore proved the following module isomorphisms:
\begin{equation} \label{ModuleIsok=4}
\affine{\mathcal{L}}^{\brac{4}}_0 \oplus \affine{\mathcal{L}}^{\brac{4}}_4 \cong \mathcal{L}^{\brac{4}}_0 \cong \affine{\mathcal{V}}^{\brac{1}}_{\pair{0}{0}} \qquad \text{and} \qquad \brac{\affine{\mathcal{L}}^{\brac{4}}_2}^{\oplus 2} \cong \mathcal{L}^{\brac{4}}_2 \cong \affine{\mathcal{V}}^{\brac{1}}_{\pair{1}{0}} \oplus \affine{\mathcal{V}}^{\brac{1}}_{\pair{0}{1}},
\end{equation}
and thus the corresponding character identities.  As noted in \secref{secExamplek=4}, monodromy charge considerations prevent us from constructing a similar isomorphism for $\mathcal{L}^{\brac{4}}_1$.

We remark that these are isomorphisms of $L_0$-graded modules, or more precisely, of $H_0$-graded modules where the $H_0$-grading on the $\func{\affine{\alg{sl}}}{3}_1$-modules is determined by (\ref{Isok=4}).  This grading on the $\func{\affine{\alg{sl}}}{3}_1$-modules is strictly coarser than the usual one by $\func{\alg{sl}}{3}$-weight, but the usual grading may be recovered by keeping track of the $\phi^{\brac{2}}_0$-eigenvalues on the $\alg{A}_4$-modules.  The isomorphisms (\ref{ModuleIsok=4}) of $L_0$-graded modules thus imply identities of the \emph{specialised} characters ($\tr q^{L_0}$) only.  To obtain an identity involving the full $\func{\affine{\alg{sl}}}{3}_1$-characters (distinguishing the conjugate modules), we would have to incorporate the $\phi^{\brac{2}}_0$-eigenvalues in the $\alg{A}_4$-grading.  We will not pursue this subtlety here.

\section{Module Bases and Characters} \label{secCharacters}

\subsection{The $\alg{A}_1$ Character} \label{secChark=1}

When $k = 1$, we saw in \secref{secSingVectsEx} that the \gcr{} with $\gamma = 2$ was sufficient to demonstrate that the singular vectors vanish.  We claim that to determine the character of the (single) irreducible $\alg{A}_1$-module, we only need to consider the $\gamma = 1$ identity:
\begin{equation} \label{eqnGCR1k=1}
\sum_{\ell = 0}^{\infty} \binom{\ell - \tfrac{1}{2}}{\ell} \sqbrac{\phi^{\brac{m}}_{r-\ell} \phi^{\brac{n}}_{s+\ell} - \brac{-1}^{m+n} \phi^{\brac{n}}_{s-1/2-\ell} \phi^{\brac{m}}_{r+1/2+\ell}} \gcreq{1} \delta_{m+n,1} \delta_{r+s,0}
\end{equation}
(note $\mathcal{S}_{m,n} = \id$ for all $m,n$).

Consider then a given quadratic term $\phi^{\brac{m}}_r \phi^{\brac{n}}_s$, for which we define the \emph{distance-$1$ difference} to be $r-s$.  \eqnref{eqnGCR1k=1} then allows us to replace this quadratic term, whenever it appears, by a linear combination of terms whose distance-$1$ differences are $r-s - 2 \ell$ (for $\ell \geqslant 1$) and $s-r-1 - 2 \ell$ (for $\ell \geqslant 0$), as well as a possible constant term.  This replacement results in a strict decrease in the distance-$1$ difference when $r-s > \tfrac{-1}{2}$, hence we find that the vector space of all such quadratic terms (and constants) is spanned by those which satisfy $r-s \leqslant \tfrac{-1}{2}$ (and constants).

We can sharpen this conclusion by considering \eqnref{eqnGCR1k=1} at the critical difference $r-s = \tfrac{-1}{2}$:
\begin{equation}
\sum_{\ell = 0}^{\infty} \binom{\ell - \tfrac{1}{2}}{\ell} \sqbrac{\phi^{\brac{m}}_{s-1/2-\ell} \phi^{\brac{n}}_{s+\ell} - \brac{-1}^{m+n} \phi^{\brac{n}}_{s-1/2-\ell} \phi^{\brac{m}}_{s+\ell}} \gcreq{1} \delta_{m+n,1} \delta_{s,1/4}.
\end{equation}
When $m=n$, this is vacuous (both sides vanish identically), but for $m \neq n$, say $m=0$ and $n=1$, it allows us to express $\phi^{\brac{0}}_{s-1/2} \phi^{\brac{1}}_s$ as a linear combination of $\phi^{\brac{1}}_{s-1/2} \phi^{\brac{0}}_s$, other quadratic terms of strictly lower distance-$1$ difference, and a constant.  Thus, our spanning set for the vector space of quadratic terms may be culled slightly by imposing the constraint that if $r-s = \tfrac{-1}{2}$, then we require that $m \geqslant n$.

Consider now a state of the form
\begin{equation}
\phi^{\brac{m_1}}_{r_1} \cdots \phi^{\brac{m_{\ell}}}_{r_{\ell}} \ket{0}.
\end{equation}
The monodromy charge of the vacuum is zero, so we must have $r_{\ell} \in \ZZ - \tfrac{1}{4}$.  As acting with a $\phi$-mode changes the monodromy charge by $k/2 = \tfrac{1}{2}$, $r_i \in \ZZ - \tfrac{1}{4}$ when $\ell - i$ is even and $r_i \in \ZZ + \tfrac{1}{4}$ when $\ell - i$ is odd.  Similarly, the $\func{\alg{u}}{1}$-charge $\theta^{\pair{m_{\ell}}{0}}_0$ of the vacuum also vanishes, hence $r_{\ell} \leqslant \tfrac{-1}{4}$.

We have just seen that any quadratic term $\phi^{\brac{m}}_r \phi^{\brac{n}}_s$ with $r-s > \tfrac{-1}{2}$, or $r-s = \tfrac{-1}{2}$ and $m < n$ may be expressed via the \gcr{} with $\gamma = 1$ as a linear combination of those desired terms with $r-s < \tfrac{-1}{2}$ or $r-s = \tfrac{-1}{2}$ and $m \geqslant n$ (and constants).  We can extend this result to states of the vacuum module considered above, so that these constraints apply (simultaneously) to every pair of consecutive modes occurring therein\footnote{This extension is not entirely trivial, but for brevity we will omit it.  We will prove this result in much greater generality in \cite{RidSU207}, and note that the analogous result for the minimal models will also be presented in \cite{RidMin06}.}.  We have therefore the following result.

\begin{proposition} \label{propSpanningSetk=1}
The irreducible $\alg{A}_1$-\hwm{} $\mathcal{L}^{\brac{1}}_0$ is spanned by states of the form
\begin{equation}
\phi^{\brac{m_1}}_{r_1} \cdots \phi^{\brac{m_{\ell}}}_{r_{\ell}} \ket{0} \qquad \text{($\ell \geqslant 0$, $m_i \in \set{0,1}$, $r_{\ell - 2i} \in \ZZ - \tfrac{1}{4}$, $r_{\ell - 2i - 1} \in \ZZ + \tfrac{1}{4}$)},
\end{equation}
where $r_{\ell} \leqslant \tfrac{-1}{4}$, $r_i - r_{i+1} \leqslant \tfrac{-1}{2}$, and $r_i - r_{i+1} = \tfrac{-1}{2}$ implies $m_i \geqslant m_{i+1}$.
\end{proposition}


We claim that the spanning set given in \propref{propSpanningSetk=1} is actually a basis.  To prove this directly, we would have to verify that this set is linearly independent, a difficult task.  However, this can be rigorously confirmed by computing the module character corresponding to the spanning set of \propref{propSpanningSetk=1}, and comparing it to the sum of the characters of the affine modules $\affine{\mathcal{L}}^{\brac{0}}_0$ and $\affine{\mathcal{L}}^{\brac{0}}_1$ (as in \corref{corModuleIso}).

To this end then, let us work out the combinatorics corresponding to this proposed basis.  To each basis state $\phi^{\brac{m_1}}_{r_1} \cdots \phi^{\brac{m_{\ell}}}_{r_{\ell}} \ket{0}$, we associate a partition\footnote{We intend to follow the standard practice \cite{AndThe76} of denoting partitions by $\lambda$, despite our previous usage of this symbol for weights.  For the remainder of this section we shall use $\mu$ for weights, trusting that our use of $\lambda$ will not cause any confusion.} $\lambda$ by 
\begin{equation}
\lambda_i = -r_i + \tfrac{3}{4} - \tfrac{1}{2} \brac{\ell - i}.
\end{equation}
It is easy to check that $\lambda$ is indeed a partition, and that if we ``colour'' its parts by the $m_i \in \set{0,1}$, then the only condition on the coloured partitions is that if $\lambda_i = \lambda_{i+1}$, then $m_i \geqslant m_{i+1}$.  That is, if any parts of the partition coincide, those that are coloured $1$ must occur to the left of those that are coloured $0$.

Given such a coloured partition $\lambda$, we can split it into two ``monochromatic'' subpartitions $\lambda^0$ and $\lambda^1$ (whose parts are all coloured $0$ or $1$ respectively).  Conversely, given any two partitions, we can put the parts together to make a composite partition, reordering the component parts to satisfy the partition condition.  If we colour the parts so as to distinguish from which of the original two partitions the part derived, then the only ambiguity in constructing the composite partition is in deciding the order when a part from one of the constituent partitions coincides with a part from the other.  This ambiguity is resolved by requiring that when the parts coincide, the parts of one colour (call it $1$) are ordered to the left of those of the other colour ($0$ say).

We therefore have a bijective correspondence between the coloured partitions corresponding to our proposed basis and pairs of ordinary (monochromatic) partitions.  Specifically, we have proved that
\begin{equation}
\begin{Bmatrix}
\text{coloured partitions} \\
\text{of length $\ell$}
\end{Bmatrix}
\quad \cong \quad \bigoplus_{\ell_0 + \ell_1 = \ell} 
\begin{Bmatrix}
\text{partitions} \\
\text{of length $\ell_0$}
\end{Bmatrix}
\times
\begin{Bmatrix}
\text{partitions} \\
\text{of length $\ell_1$}
\end{Bmatrix}
.
\end{equation}
As the conformal dimensions of our basis states are related to the weights $\abs{\lambda} = \sum_i \lambda_i$ of the associated partitions by 
\begin{equation}
h = \abs{\lambda} + \tfrac{1}{4} \ell \brac{\ell - 4},
\end{equation}
and the $\func{\alg{sl}}{2}$-weights are related to the colourings by
\begin{equation}
w = \ell - 2 \sum_i m_i = \ell_0 - \ell_1,
\end{equation}
the character of the $\alg{A}_1$-vacuum module is
\begin{align}
\func{\chi^{\brac{1}}_0}{q ; z} &= \sum_{\text{basis}} q^h z^w = \sum_{\ell_0 , \ell_1 = 0}^{\infty} \ \sideset{}{'} \sum_{\lambda^0 , \lambda^1} q^{\abs{\lambda^0} + \abs{\lambda^1} + \tfrac{1}{4} \brac{\ell_0 + \ell_1}^2 - \ell_0 - \ell_1} z^{\ell_0 - \ell_1} \notag \\
&= \sum_{\ell_0 , \ell_1 = 0}^{\infty} \frac{q^{\brac{\ell_0 + \ell_1}^2 / 4}}{\qfact{q}{\ell_0} \qfact{q}{\ell_1}} z^{\ell_0 - \ell_1} = \sum_{\ell = 0}^{\infty} \sum_{m=0}^{\ell} \frac{q^{\ell^2 / 4}}{\qfact{q}{m} \qfact{q}{\ell-m}} z^{\ell - 2m},
\end{align}
where $\qfact{q}{m} = \prod_{j=1}^m \brac{1 - q^j}$ as usual, and the primed summation indicates that the partition $\lambda^i$ has length $\ell_i$.  Here, we have used the well known partition identity \cite[Eq.\ 2.2.5]{AndThe76}
\begin{equation}
\sum_{\ell = 0}^{\infty} \sideset{}{'} \sum_{\lambda} q^{\abs{\lambda}} z^{\ell} = \prod_{j=1}^{\infty} \frac{1}{1 - z q^j} = \sum_{\ell = 0}^{\infty} \frac{q^{\ell}}{\qfact{q}{\ell}} z^{\ell}.
\end{equation}
The affine characters are obtained by restricting $\ell$ to be even or odd, by \lemref{lemTwoPhi=Aff}.  They then take the form
\begin{equation}
\begin{split}
\func{\affine{\chi}^{\brac{1}}_0}{q ; z} &= \sum_{n=0}^{\infty} \sum_{m=0}^{2n} \frac{q^{n^2}}{\qfact{q}{m} \qfact{q}{2n-m}} z^{2 \brac{n-m}} \\
\text{and} \qquad \func{\affine{\chi}^{\brac{1}}_1}{q ; z} &= z q^{1/4} \sum_{n=0}^{\infty} \sum_{m=0}^{2n+1} \frac{q^{n^2 + n}}{\qfact{q}{m} \qfact{q}{2n+1-m}} z^{2 \brac{n-m}},
\end{split}
\end{equation}
where the subscript denotes the highest weight.

It is shown in \cite{KedSum95} that these formulae (ignoring the $z$-dependence), agree with the usual bosonic character formulae, completing our proof that the spanning set given in \propref{propSpanningSetk=1} is indeed a basis as claimed.

\newpage

\begin{theorem} \label{thmBasisk=1}
A basis for the (irreducible) $\alg{A}_1$-vacuum module $\mathcal{L}^{\brac{1}}_0$ is given by the states:
\begin{equation}
\phi^{\brac{m_1}}_{r_1} \cdots \phi^{\brac{m_{\ell}}}_{r_{\ell}} \ket{0} \qquad \text{($\ell \geqslant 0$, $m_i \in \set{0,1}$, $r_{\ell - 2i} \in \ZZ - \tfrac{1}{4}$, $r_{\ell - 2i - 1} \in \ZZ + \tfrac{1}{4}$)},
\end{equation}
where $r_{\ell} \leqslant \tfrac{-1}{4}$, $r_i - r_{i+1} \leqslant \tfrac{-1}{2}$, and $r_i - r_{i+1} = \tfrac{-1}{2}$ implies $m_i \geqslant m_{i+1}$.  The character of this module is
\begin{equation}
\func{\chi^{\brac{1}}_0}{q ; z} = \sum_{\ell_0 , \ell_1 = 0}^{\infty} \frac{q^{\brac{\ell_0 + \ell_1}^2 / 4}}{\qfact{q}{\ell_0} \qfact{q}{\ell_1}} z^{\ell_0 - \ell_1}.
\end{equation}
\end{theorem}

\noindent We mention that the character formula we have derived agrees with that obtained in \cite{BouSpi94} from a proposed ``spinon basis'', even though our basis is quite different.  Our \thmref{thmBasisk=1} may therefore be viewed as an elementary proof of their results (for $k=1$).

\subsection{The $\alg{A}_2$ Characters} \label{secChark=2}

The defining \opes{} when $k=2$ have already been given in \secref{secExamplek=2}, where it was noted (\eqnref{eqnSuperCommk=2}) that the \gcrs{} with $\gamma = 1$ are supercommutation relations.  We claim that considering these supercommutation relations is sufficient for computing the characters of the $\alg{A}_2$-modules.  We present a proof of this claim below which is entirely analogous to that given for $k=1$.

As in \secref{secChark=1}, the distance-$1$ difference $r-s$ of a quadratic term $\phi^{\brac{m}}_r \phi^{\brac{n}}_s$ may be strictly reduced by using the supercommutation relation if $r > s$ (we swap the modes).  The vector space of quadratic terms is thus spanned by those with distance-$1$ difference at most $0$.  When $r = s$, we may swap a quadratic term with $m < n$ for one with $m > n$, and those with $m = n$ are seen to be equivalent to constants.  The analogue of \propref{propSpanningSetk=1} is therefore:

\begin{proposition} \label{propSpanningSetk=2}
The irreducible $\alg{A}_2$-\hwm{} $\mathcal{L}^{\brac{2}}_{\mu}$ ($\mu \in \set{0, 1}$) is spanned by states of the form
\begin{equation}
\phi^{\brac{m_1}}_{r_1} \cdots \phi^{\brac{m_{\ell}}}_{r_{\ell}} \ket{\psi_{\mu}} \qquad \text{($\ell \geqslant 0$, $m_i \in \set{0,1,2}$, $r_i \in \ZZ + \brac{\mu - 1} / 2$),}
\end{equation}
where $r_{\ell} \leqslant \theta^{\pair{m_{\ell}}{0}}_{\mu} - \tfrac{1}{2}$, $r_i \leqslant r_{i+1}$, and $r_i = r_{i+1}$ implies $m_i > m_{i+1}$.
\end{proposition}

Let us therefore associate a partition $\lambda$ to each basis state $\phi^{\brac{m_1}}_{r_1} \cdots \phi^{\brac{m_{\ell}}}_{r_{\ell}} \ket{\psi_{\mu}}$ by
\begin{equation}
\lambda_i = -r_i + \brac{\mu + 1} / 2.
\end{equation}
Since $\theta^{\pair{m_{\ell}}{0}}_{\mu} \leqslant \mu / 2$, it is easy to check that $\lambda$ is a genuine partition which, when coloured by the $m_i$, satisfies the ordering condition that if $\lambda_i = \lambda_{i+1}$, then $m_i > m_{i+1}$.

We therefore decompose such a partition $\lambda$ into its three monochromatic subpartitions $\lambda^0$, $\lambda^1$ and $\lambda^2$, noting that the ordering condition implies that each $\lambda^m$ is a partition into distinct parts.  Conversely, given three partitions into distinct parts, we can uniquely merge them into a coloured partition satisfying our ordering condition.  This bijection of partitions leads to an expression for the character of the vacuum module ($\mu = 0$):
\begin{equation}
\func{\chi^{\brac{2}}_0}{q ; z} = \sum_{\ell_0, \ell_1, \ell_2 = 0}^{\infty} \ \sideset{}{'} \sum_{\lambda^0, \lambda^1, \lambda^2} q^{\abs{\lambda^0} + \abs{\lambda^1} + \abs{\lambda^2} - \tfrac{1}{2} \brac{\ell_0 + \ell_1 + \ell_2}} z^{2 \brac{\ell_0 - \ell_2}}.
\end{equation}
The prime on the summation indicates that we are summing over partitions $\lambda^m$ of length $\ell_m$ into distinct parts, and we have noted that the conformal dimension of our state and its $\func{\alg{sl}}{2}$-weight are given by
\begin{equation}
h_{\mu} + \abs{\lambda} - \ell \brac{\mu + 1} / 2 \qquad \text{and} \qquad 2 \Bigl( \ell - \sum_i m_i \Bigr) = 2 \brac{\ell_0 - \ell_2},
\end{equation}
respectively.  The generating function for partitions into distinct parts, \cite[Eq.\ 2.2.6]{AndThe76}
\begin{equation} \label{eqnEulerGenFunc}
\sum_{\ell = 0}^{\infty} \sideset{}{'} \sum_{\lambda} q^{\abs{\lambda}} z^{\ell} = \prod_{j=1}^{\infty} \brac{1 + z q^j} = \sum_{\ell = 0}^{\infty} \frac{q^{\ell \brac{\ell + 1} / 2}}{\qfact{q}{\ell}} z^{\ell},
\end{equation}
then gives
\begin{equation}
\func{\chi^{\brac{2}}_0}{q ; z} = \sum_{\ell_0, \ell_1, \ell_2 = 0}^{\infty} \frac{q^{\brac{\ell_0^2 + \ell_1^2 + \ell_2^2} / 2}}{\qfact{q}{\ell_0} \qfact{q}{\ell_1} \qfact{q}{\ell_2}} z^{2 \brac{\ell_0 - \ell_2}}.
\end{equation}
Dropping the $z$-dependence, this is exactly the character of the fermionic vacuum module cubed, which we noted in \secref{secExamplesRevisited} is the character of $\mathcal{L}^{\brac{2}}_0$.  This proves that our spanning set is indeed a basis (for the vacuum module), because it gives the correct character.

To compute the other character ($\mu = 1$), there is one further constraint, because $\theta^{\pair{m_{\ell}}{0}}_{\mu} \neq \mu / 2$ in general (but equality always holds when $\mu = 0$).  We therefore need to impose the condition $\lambda_{\ell} \geqslant \tfrac{3}{2} - \theta^{\pair{m_{\ell}}{0}}_{\mu}$.  This bound is $2$ or $1$, according as to whether $m_{\ell}$ is $0$ or not.  It follows that the monochromatic subpartition $\lambda^0$ must be further constrained not to have a part equal to $1$.  Such subpartitions are in bijective correspondence with partitions into distinct parts (subtract $1$ from each part of the former), so this additional condition is easily accommodated.  The final result for the character is then
\begin{equation}
\func{\chi^{\brac{2}}_1}{q ; z} = z q^{3 / 16} \sum_{\ell_0, \ell_1, \ell_2 = 0}^{\infty} \frac{q^{\brac{\ell_0^2 + \ell_1^2 + \ell_2^2} / 2 + \brac{\ell_0 - \ell_1 - \ell_2} / 2}}{\qfact{q}{\ell_0} \qfact{q}{\ell_1} \qfact{q}{\ell_2}} z^{2 \brac{\ell_0 - \ell_2}}.
\end{equation}

If we substitute $z = q^{-1}$ into \eqnref{eqnEulerGenFunc}, we derive that
\begin{equation}
\sum_{\ell = 0}^{\infty} \frac{q^{\ell \brac{\ell - 1} / 2}}{\qfact{q}{\ell}} = \prod_{j=0}^{\infty} \brac{1 + q^j} = 2 \prod_{j=1}^{\infty} \brac{1 + q^j} = 2 \sum_{\ell = 0}^{\infty} \frac{q^{\ell \brac{\ell + 1} / 2}}{\qfact{q}{\ell}}.
\end{equation}
It now follows that
\begin{equation}
\func{\chi^{\brac{2}}_1}{q ; 1} = \tfrac{1}{2} \sqbrac{q^{1 / 16} \sum_{\ell = 0}^{\infty} \frac{q^{\ell \brac{\ell - 1} / 2}}{\qfact{q}{\ell}}}^3,
\end{equation}
that is, half the cube of the fermionic module whose \hws{} has conformal dimension $1/16$.  As this is indeed the correct character of $\mathcal{L}^{\brac{2}}_1$ (\secref{secExamplesRevisited}), we may again conclude that our spanning set is in fact a basis for this module.  We have therefore completed the proof of the following theorem:

\newpage

\begin{theorem} \label{thmBasisk=2}
A basis for the (irreducible) $\alg{A}_1$-vacuum module $\mathcal{L}^{\brac{2}}_0$ is given by the states:
\begin{equation}
\phi^{\brac{m_1}}_{r_1} \cdots \phi^{\brac{m_{\ell}}}_{r_{\ell}} \ket{0} \qquad \text{($\ell \geqslant 0$, $m_i \in \set{0,1,2}$, $r_i \in \ZZ - \tfrac{1}{2}$),}
\end{equation}
where $r_{\ell} \leqslant \tfrac{-1}{2}$, $r_i \leqslant r_{i+1}$, and $r_i = r_{i+1}$ implies $m_i > m_{i+1}$.  The corresponding character is
\begin{equation}
\func{\chi^{\brac{2}}_0}{q ; z} = \sum_{\ell_0, \ell_1, \ell_2 = 0}^{\infty} \frac{q^{\brac{\ell_0^2 + \ell_1^2 + \ell_2^2} / 2}}{\qfact{q}{\ell_0} \qfact{q}{\ell_1} \qfact{q}{\ell_2}} z^{2 \brac{\ell_0 - \ell_2}}.
\end{equation}
A basis for the (irreducible) $\alg{A}_2$-module $\mathcal{L}^{\brac{2}}_1$ is given by the states
\begin{equation}
\phi^{\brac{m_1}}_{r_1} \cdots \phi^{\brac{m_{\ell}}}_{r_{\ell}} \ket{\psi_1} \qquad \text{($\ell \geqslant 0$, $m_i \in \set{0,1,2}$, $r_i \in \ZZ$),}
\end{equation}
where $r_{\ell} \leqslant 0$, $r_{\ell} = 0$ implies $m_{\ell} > 0$, $r_i \leqslant r_{i+1}$, and $r_i = r_{i+1}$ implies $m_i > m_{i+1}$.  The corresponding character is
\begin{equation}
\func{\chi^{\brac{2}}_1}{q ; z} = z q^{3 / 16} \sum_{\ell_0, \ell_1, \ell_2 = 0}^{\infty} \frac{q^{\brac{\ell_0^2 + \ell_1^2 + \ell_2^2} / 2 + \brac{\ell_0 - \ell_1 - \ell_2} / 2}}{\qfact{q}{\ell_0} \qfact{q}{\ell_1} \qfact{q}{\ell_2}} z^{2 \brac{\ell_0 - \ell_2}}.
\end{equation}
\end{theorem}


\newpage

\begin{thebibliography}{10}

\bibitem{KniCur84}
V~Knizhnik and A~Zamolodchikov.
\newblock {Current Algebra and Wess-Zumino Model in Two Dimensions}.
\newblock {\em Nucl. Phys.}, B247:83--103, 1984.

\bibitem{WitNon84}
E~Witten.
\newblock {Non-abelian Bosonization in Two Dimensions}.
\newblock {\em Comm. Math. Phys.}, 92:455--472, 1984.

\bibitem{GepStr86}
D~Gepner and E~Witten.
\newblock {String Theory on Group Manifolds}.
\newblock {\em Nucl. Phys.}, B278:493--549, 1986.

\bibitem{FelSpe88}
G~Felder, K~Gaw\c{e}dzki, and A~Kupiainen.
\newblock {The Spectrum of Wess-Zumino-Witten Models}.
\newblock {\em Nucl. Phys.}, B299:355--366, 1988.

\bibitem{DiFCon97}
P~Di Francesco, P~Mathieu, and D~S\'{e}n\'{e}chal.
\newblock {\em {Conformal Field Theory}}.
\newblock Graduate Texts in Contemporary Physics. Springer-Verlag, New York,
  1997.

\bibitem{BerSpi94}
D~Bernard, V~Pasquier, and D~Serban.
\newblock {Spinons in Conformal Field Theory}.
\newblock {\em Nucl. Phys.}, B428:612--628, 1994.
\newblock \texttt{arXiv:hep-th/9404050}.

\bibitem{BouSpi94}
P~Bouwknegt, A~Ludwig, and K~Schoutens.
\newblock {Spinon Bases, Yangian Symmetry and Fermionic Representations of
  Virasoro Characters in Conformal Field Theory}.
\newblock {\em Phys. Lett.}, B338:448--456, 1994.
\newblock \texttt{arXiv:hep-th/9406020}.

\bibitem{BouSpi95}
P~Bouwknegt, A~Ludwig, and K~Schoutens.
\newblock {Spinon Basis for Higher Level $\mathsf{SU} \left( 2 \right)$ WZW
  Models}.
\newblock {\em Phys. Lett.}, B359:304--312, 1995.
\newblock \texttt{arXiv:hep-th/9412108}.

\bibitem{FeiMon05}
B~Feigin, M~Jimbo, T~Miwa, E~Mukhin, and Y~Takeyama.
\newblock {A Monomial Basis for the Virasoro Minimal Series $M \left( p , p'
  \right)$: The Case $1 < p'/p < 2$}.
\newblock {\em Comm. Math. Phys.}, 257:395--423, 2005.
\newblock \texttt{arXiv:math.QA/0405468}.

\bibitem{JacQua06}
P~Jacob and P~Mathieu.
\newblock {A Quasi-Particle Description of the $M \left( 3 , p \right)$
  Models}.
\newblock {\em Nucl. Phys.}, B733:205--232, 2006.
\newblock \texttt{arXiv:hep-th/0506074}.

\bibitem{JacEmb06}
P~Jacob and P~Mathieu.
\newblock {Embedding of Bases: From the $M \left( 2 , 2 \kappa + 1 \right)$ to the $M
  \left( 3 , 4 \kappa + 2 - \delta \right)$ Models}.
\newblock {\em Phys. Lett.}, B635:350--354, 2006.
\newblock \texttt{arXiv:hep-th/0511040}.

\bibitem{RidMin06}
P~Mathieu and D~Ridout.
\newblock {The Extended Algebra of the Minimal Models}.
\newblock (In preparation).

\bibitem{GodKac85}
P~Goddard and D~Olive.
\newblock {Kac-Moody Algebras, Conformal Symmetry and Critical Exponents}.
\newblock {\em Nucl. Phys.}, B257:226--252, 1985.

\bibitem{BaiAcc86}
F~Bais and A~Taormina.
\newblock {Accidental Degeneracies in String Compactification}.
\newblock {\em Phys. Lett.}, B181:87--90, 1986.

\bibitem{RidSU207}
P~Mathieu and D~Ridout.
\newblock {Characters of the Extended Algebra of the $SU \left( 2 \right)$
  Wess-Zumino-Witten Models}.
\newblock (In preparation).

\bibitem{KacInf90}
V~Kac.
\newblock {\em {Infinite-Dimensional Lie Algebras}}.
\newblock Cambridge University Press, Cambridge, 1990.

\bibitem{ThiOPE91}
K~Thielemans.
\newblock {A \textsc{Mathematica} Package for Computing Operator Product
  Expansions}.
\newblock {\em Int. J. Mod. Phys.}, C2:787--798, 1991.
\newblock Package available at
  \texttt{http://www.hammersmithimanet.com/$\sim$kris/string/}.

\bibitem{AndThe76}
G~Andrews.
\newblock {\em {The Theory of Partitions}}, volume~2 of {\em Encyclopedia of
  Mathematics and its Applications}.
\newblock Addison-Wesley, Reading, 1976.

\bibitem{KedSum95}
R~Kedem, B~McCoy, and E~Melzer.
\newblock {The Sums of Rogers, Schur and Ramanujan and the Bose-Fermi
  Correspondence in $\left( 1+1 \right)$-Dimensional Quantum Field Theory}.
\newblock In P~Bouwknegt \emph{et al}, editor, {\em Recent Progress in
  Statistical Mechanics and Quantum Field Theory}, pages 195--219. World
  Scientific, New Jersey, 1995.
\newblock \texttt{arXiv:hep-th/9304056}.

\end{thebibliography}
\end{document}